\documentclass[reprint, superscriptaddress,aps,pra,amsmath,amssymb
]{revtex4-2}
\usepackage{bbm}
\usepackage[british]{babel}
\usepackage{graphicx}
\usepackage{dcolumn}
\usepackage{bm}
\usepackage{hyperref}
\usepackage[title]{appendix}
\usepackage{dsfont}
\usepackage{tipa}
\usepackage[usenames,dvipsnames]{color}
\usepackage{xfrac}
\usepackage{amsmath}
\usepackage{leftindex}
\usepackage{graphicx}
\usepackage{caption}
\usepackage[subrefformat=parens,labelformat=parens]{subcaption}
\usepackage{newtx}
\usepackage{comment}
\usepackage{listings}
\usepackage{color}
\usepackage{mathtools}

\definecolor{dkgreen}{rgb}{0,0.6,0}
\definecolor{gray}{rgb}{0.5,0.5,0.5}
\definecolor{mauve}{rgb}{0.58,0,0.82}

\providecommand{\ket}[1]{| #1{\rangle}}

\providecommand{\bra}[1]{\langle #1|}

\providecommand{\sprod}[2]{\langle #1|#2\rangle}

\DeclarePairedDelimiter\abs{\lvert}{\rvert}

\makeatletter

\renewcommand\p@subfigure{\thefigure}
\makeatother

\newcommand{\red}[1]{#1}

\begin{document}


\title{Looking down the rabbit hole: Towards quantum optimal estimation of surface roughness}

\author{Quentin Muller}
\affiliation{School of Mathematical Sciences and Centre for the Mathematics and Theoretical Physics of Quantum Non-Equilibrium Systems, University of Nottingham, University Park, Nottingham NG7 2RD, UK}

\author{Tommaso Tufarelli}
\affiliation{Algorithmiq, Via della Chiusa 15,
20123 Milano, Italia}

\author{M\u{a}d\u{a}lin Gu\c{t}\u{a}}
\affiliation{School of Mathematical Sciences and Centre for the Mathematics and Theoretical Physics of Quantum Non-Equilibrium Systems, University of Nottingham, University Park, Nottingham NG7 2RD, UK}

\author{Katherine Inzani}
\affiliation{School of Chemistry, University of Nottingham, University Park, Nottingham NG7 2RD, UK}

\author{Samanta Piano}
\affiliation{Manufacturing Metrology Team, Faculty of Engineering, University of Nottingham, Nottingham NG7 2RD, UK}

\author{Gerardo Adesso}
\email{Corresponding author: gerardo.adesso@nottingham.ac.uk}
\affiliation{School of Mathematical Sciences and Centre for the Mathematics and Theoretical Physics of Quantum Non-Equilibrium Systems, University of Nottingham, University Park, Nottingham NG7 2RD, UK}


\begin{abstract}
Surface roughness is an important quantity to many engineering and precision manufacturing disciplines. In this paper we investigate the problem of estimating the root-mean-square roughness of a sample by passive linear optics. 
\red{By adopting quantum parameter estimation methods, we determine the ultimate
precision limits for estimating spatial moments of a general three-dimensional distribution of incoherent point
sources in the sub-diffraction regime. Specializing this result to the axial profile,} we show that the information
on the first moment (mean height) and standard deviation (roughness) is bounded by a constant. While classical imaging techniques fail to achieve this bound, a quantum inspired imaging technique based on spatial mode demultiplexing is proven to be optimal for estimating the axial standard deviation. \red{This provides a powerful and experimentally accessible route} to measuring roughness of nearly smooth surface patches beyond the diffraction limit. 
\end{abstract}


\maketitle

\section{Introduction}
Surface roughness plays a central role in diverse scientific and industrial contexts. In precision manufacturing, surface finish directly affects mechanical performance, wear, and lubrication properties ~\cite{Bhushan1999Tribology}. In aerospace and automotive engineering, surface roughness influences friction, drag, and fatigue life~\cite{Stout1993Surface}. In optics and photonics, nanometre-scale roughness of mirrors, gratings, or coatings determines scattering losses and limits device performance~\cite{Elson1995Optics}. In semiconductor fabrication and nanotechnology, roughness control is critical for layer deposition and pattern transfer~\cite{Zhou2002Semiconductor}. Roughness is also a critical quality metric in additive manufacturing and precision metrology, where it impacts dimensional tolerances and functional compliance~\cite{Leach2013Book}. Given this broad relevance, accurate measurement and characterization of surface roughness---and understanding the fundamental limits of such estimation---are of great fundamental and applied interest.

Rough surfaces can be thought of as ``miniature mountain ranges,'' with peaks and troughs forming a height distribution over the surface. A variety of metrics exist to characterize roughness~\cite{ISOstandard}, but in this work we focus on the \emph{root-mean-square height variation} (the standard deviation of the surface height profile). We denote this roughness parameter by $\sigma$, the square root of the second central moment of the height distribution:
\begin{equation}\label{sigma}
\sigma = \sqrt{\theta_2 - \theta_1^2}\,,
\end{equation}
where 
 \begin{equation}
 \label{object plane moment}
     \theta_j=\int z^j dF(z)
 \end{equation}
 is the $j^{\text{th}}$ moment of the height distribution $F(z)$; in particular, $\theta_1$ is the mean height and $\sigma^2=\theta_2-\theta_1^2$ is the variance of the surface height profile.


\begin{figure}[ht]
\centering
\includegraphics[width=0.2\textwidth]{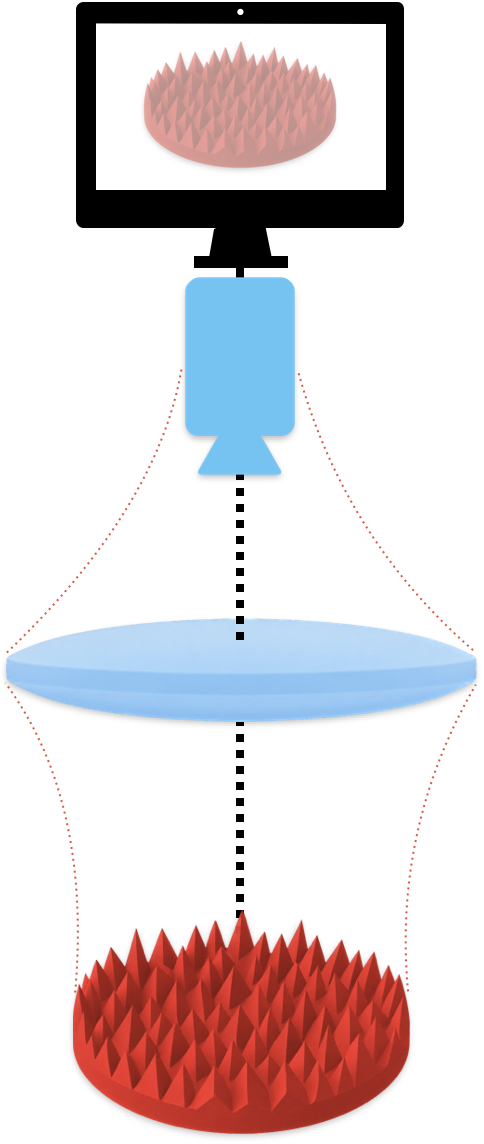}
\caption{A schematic displaying the focusing of a small profile of a rough surface through a lens. The image of the rough surface can then be attained via different measurement schemes.}
\label{MountainRange}
\end{figure}

Techniques for measuring the surface roughness can be separated into contact and non-contact methods \cite{Allardyce:87}. A common contact technique is the use of a stylus profilometer --- a diamond tipped stylus that is dragged across a rough surface to produce a varying voltage proportional to the height of the surface contour. However, such a contact technique may damage the surface and only yields $1D$ information. As such, we turn our attention instead to non-contact imaging using optical sensors. In such an attempt to estimate the surface distribution, coherent light, usually in the form of a monochromatic plane wave \cite{PhysRevE.71.036606}, is made incident upon the landscape, and the reflected light is then analyzed. 
However, as the surface roughness becomes smaller and the relative heights and radial displacements of the peaks and troughs dip below the Rayleigh limit \cite{Rayleigh}, diffraction effects resulting from the focusing of the emitted light through some aperture or lens take place, leading to loss of information at the image plane. It becomes therefore an interesting metrological problem to devise more efficient ways of passively collecting and processing light emitted from rough surfaces in the sub-diffraction limit, in order to improve both the angular and axial resolution for reconstructing relevant surface parameters.

Here we address this problem using methods from {\em quantum metrology}. Our analysis is inspired by the recent body of work that reformulated the task of resolving spatially displaced point sources as a quantum parameter estimation problem \cite{PhysRevX.6.031033,Tsang_2017,Rehacek:17,Ang17,PhysRevA.99.012305,PhysRevLett.122.140505,fiderer2021general,lvovsky2026passive}. These groundbreaking papers not only surpassed the centuries-old Rayleigh limit through a quantum-inspired measurement technique known as spatial mode demultiplexing (SPADE), but also placed the notion of resolution on a more rigorous footing by connecting it to the minimum achievable error in estimating the defining parameters of a source distribution; namely, the geometric moments. Much of the work carried out since in this realm has used either {\it radially} distributed light sources \cite{tsang2019resolving} (i.e., displaced within a plane parallel to the image plane) or at most two point sources distributed {\it axially} \cite{PhysRevLett.122.140505,Zhou:19} (i.e., displaced along a line perpendicular to the image plane). 

In this paper we expand on the previously considered cases by treating the problem of an \emph{arbitrary} number of discrete \emph{axially} distributed point sources: an optical ``rabbit hole''. \red{This problem is analyzed as a special instance of the general endeavor to characterize uncertainty in estimating light sources distributed both axially and radially as in a complete three-dimensional point cloud, for which we determine fundamental precision bounds in the sub-diffraction regime.}  In particular, complementing existing results for moments estimation in radial source distributions \cite{PhysRevA.99.013808,Tsang_2017,PhysRevResearch.1.033006}, here we compute the ultimate limits on the information available in estimating the first moment --- mean height --- and (square root of) the second central moment --- roughness --- of the axial distribution of $N$ incoherent point sources. We find that classical direct imaging techniques fail to reach these limits, and we show the quantum-inspired measurement technique SPADE (specifically decomposing output modes in the Laguerre-Gauss basis) to be optimal in fulfilling the considered estimation problem.

This paper is partitioned into five sections. In section~\ref{SectionBackground} we give an overview of relevant quantum metrology methods followed by a brief explanation of quantum imaging theory along with a description of the theoretical model used in this analysis. In section~\ref{SectionComputingQFI} we present our derivation of the fundamental bound on the information available for the estimation of root-mean-square roughness. In section~\ref{SectionResults} we show  the failure of classical direct imaging techniques to reach this bound as opposed to the success of \red{ideal SPADE measurements to achieve optimality for the task of roughness estimation, and we investigate the robustness of these findings against common experimental imperfections.} Finally we discuss our results and future outlook in section~\ref{SectionDiscussion}.

\section{Theoretical background}\label{SectionBackground}

\subsection{Quantum Metrology Theory}
\label{SectionMetrologyTheory}

For clarity we divide this subsection into two parts: the first part reviews the theory of quantum estimation for finite-dimensional parameters while the second part deals with the semi-parametric formalism \cite{Tsiatis,PhysRevX.10.031023,PhysRevResearch.1.033006}. The latter becomes useful for problems involving a very high-dimensional (and potentially infinite-dimensional) space of parameters. 

\subsubsection{Quantum parameter estimation and the Fisher information}

Quantum estimation deals with the following problem: given a quantum system, whose state \red{--- described by a density operator $\rho_{\boldsymbol{\lambda}}$ acting on the system's Hilbert space $\mathcal{H}$ ---} depends smoothly on an unknown parameter set
$\boldsymbol{\lambda}=(\lambda_1,\ldots,\lambda_n)^T\in\Lambda\subset\mathbb{R}^n$, the task is to estimate the 
parameters by measuring the system \cite{doi:10.1142/S0219749909004839}. Since different measurements may be mutually incompatible and convey different information about the parameters, the key questions are which measurements are most suitable for estimating 
$\boldsymbol{\lambda}$ and what is the optimal estimation precision.  


A measurement is described 
by \red{a positive operator-valued measure (POVM) 
$\{\Pi_{\omega}\}_{\omega\in\Omega}$, where each 
$\Pi_{\omega}$ is a positive semidefinite operator acting on the system's
Hilbert space $\mathcal{H}$} and the operators satisfy the completeness relation
$\int_{\Omega}\Pi_{\omega}\,d\omega = \mathbb{I}$, where $d\omega$ is a reference measure on $\Omega$. 
The probability density \red{of obtaining the outcome 
$\omega\in\Omega$ when measuring the system in the state $\rho_{\boldsymbol{\lambda}}$}, given the underlying parameters
$\boldsymbol{\lambda}$ of the model, is then given by the Born rule
\begin{equation}
    p_{\boldsymbol{\lambda}}(\omega)=\text{Tr}[\rho_{\boldsymbol{\lambda}}\Pi_{\omega}].
\end{equation}
The collection of all such smoothly parametrized probability density functions
constitutes a statistical model,
\begin{equation}
    \mathcal{P}
    = \{\,p_{\boldsymbol{\lambda}}(\omega) :
       \boldsymbol{\lambda}\in\Lambda\subset\mathbb{R}^n\,\},
\end{equation}
which will serve as the playing field for all statistical calculations.
Typically, a single measurement is insufficient to estimate $\boldsymbol{\lambda}$ to the desired precision. Therefore, the experimenter needs to perform repeated measurements on identically prepared systems and estimate the parameter from the collection of outcomes, by using a general method such as maximum likelihood, or more specific techniques tailored to the given model.

The {\it Cram\'er-Rao bound} (CRB) and its quantum counterpart provide a first hand answer to the question posed in the beginning of this section. Given an unbiased estimator $\boldsymbol{\hat{\lambda}}$ (i.e. $\mathbb{E}_{\boldsymbol{\lambda}} (\boldsymbol{\hat{\lambda}}) = \boldsymbol{\lambda}$), its covariance matrix is lower bounded as follows
\begin{equation}
\label{QCRB}
\begin{aligned}
     \text{Cov}_{\boldsymbol{\lambda}}[\boldsymbol{\hat{\lambda}}]
     \geq \mathcal{I}(\boldsymbol{\lambda};\{\Pi_{\omega}\})^{-1} \geq \mathcal{K}(\boldsymbol{\lambda})^{-1}\,.
    \end{aligned}
\end{equation}
The leftmost inequality tells us that the covariance matrix $\text{Cov}_{\boldsymbol{\lambda}}[\boldsymbol{\hat{\lambda}}]
     :=\mathbb{E}_{\boldsymbol{\lambda}}\left[(\hat{\boldsymbol{\lambda}}-{\boldsymbol{\lambda}})^T(\hat{\boldsymbol{\lambda}}-{\boldsymbol{\lambda}})\right]$
is bounded below by the inverse of the classical Fisher information matrix (CFIm) $\mathcal{I}(\boldsymbol{\lambda},\{\Pi_\omega\})$, whose elements are given by
\begin{equation}
\label{CFIm}
    [\mathcal{I}(\boldsymbol{\lambda};\{\Pi_{\omega}\})]_{ij}=\int \frac{\partial p_{\boldsymbol{\lambda}}(\omega)}{\partial\lambda_i}\frac{\partial p_{\boldsymbol{\lambda}}(\omega)}{\partial\lambda_j}\frac{d\omega}{p_{\boldsymbol{\lambda}}(\omega)}.
\end{equation}

To understand the CFIm 
more intuitively, we can think of the statistical model $\mathcal{P}$ as a submanifold of the space of distributions on $\Omega$, and the map $\boldsymbol{\lambda}: \Lambda\to \mathcal{P}$ as a coordinates map. The log-likelihood random variable is defined as 
$\ell_{\boldsymbol{\lambda}}(\omega) := \log p_{\boldsymbol{\lambda}}(\omega)$ and its differential $d\ell_{\boldsymbol{\lambda}}(\omega)$ is a one-form with components given by the  \emph{score functions} $S_{\boldsymbol{\lambda},i}(\omega) := \partial\ell_{\boldsymbol{\lambda}}(\omega)/\partial \lambda_i$. Given a tangent vector $\boldsymbol{v}=(v_1,\dots v_n)$, the corresponding directional score is 
$S_{\boldsymbol{\lambda},\boldsymbol{v}} = d\ell(\boldsymbol{\lambda},\omega)[\boldsymbol{v}]$, and is an element of 
$L^2(\Omega, p_{\boldsymbol{\lambda}})$ with expectation zero, $\mathbb{E}_{\boldsymbol{\lambda}}(S_{\boldsymbol{\lambda},\boldsymbol{v}}) =0$. The Fisher information can be seen as a Riemannian metric whose components in the coordinate system $\boldsymbol{\lambda}$ are
\begin{equation}
    [\mathcal{I}(\boldsymbol{\lambda};\{\Pi_{\omega}\})]_{ij}=
    \langle S_i,S_j\rangle_{\boldsymbol{\lambda} }\equiv\mathbb{E}_{\boldsymbol{\lambda}}[S_iS_j].
\end{equation}
Assuming that the CFIm is strictly positive, the score functions are linearly independent and form a basis in the co-tangent space $\mathcal{T}_{\boldsymbol{\lambda}}$ at $\boldsymbol{\lambda}$. Given this interpretation of the CFIm as a metric on the probability manifold, we can write the infinitesimal distance between two points in terms of the Fisher information metric as
\begin{equation}
    ds^2_p=d\boldsymbol{\lambda}^T\mathcal{I}(\boldsymbol{\lambda};\{\Pi_{\omega}\})d\boldsymbol{\lambda}.
\end{equation}

The rightmost inequality in \eqref{QCRB} is the {\em quantum} Cram\'{e}r-Rao bound (QCRB) which features the inverse of the quantum Fisher information matrix (QFIm) 
$ 
\mathcal{K}(\boldsymbol{\lambda})
$ and is independent of the measurement. Similarly to the CFIm, we can also think of the QFIm in a geometric fashion. Here we consider a manifold $\mathcal{S}$ consisting of density operators as a subset of the bounded operators $\mathcal{B}(\mathcal{H})$ acting on a Hilbert space $\mathcal{H}$,
\begin{equation}
    \mathcal{S}=\{\rho\in\mathcal{B}(\mathcal{H}):\rho=\rho^{\dagger}\geq0,\text{Tr}[\rho]=1\}.
\end{equation}
We can likewise define the infinitesimal distance between two nearby quantum states on the density operator manifold as,
\begin{eqnarray}
    ds^2_\rho= \max_{\{\Pi_{\omega}\}}d\boldsymbol{\lambda}^{T}\mathcal{I}(\boldsymbol{\lambda};\{\Pi_{\omega}\})d\boldsymbol{\lambda}\equiv d\boldsymbol{\lambda}^T\mathcal{K}(\boldsymbol{\lambda})d\boldsymbol{\lambda},
\end{eqnarray}~
where the QFIm $\mathcal{K}(\boldsymbol{\lambda})$ satisfies the matrix inequality $I(\boldsymbol{\lambda};\{\Pi_\omega\}) \le \mathcal{K}(\boldsymbol{\lambda})$ for all POVMs $\{\Pi_\omega\}$.

Explicitly, the QFIm can be written in terms of the {\it symmetric logarithmic
derivative} (SLD) operators $\mathcal{L}_i$, introduced by Holevo \cite{Holevo1982} as the non-commutative counterparts to the score functions $S_i$, and defined  by the Lyapunov equation
\begin{equation}\label{SLD}
   \frac{\partial\rho_{\boldsymbol{\lambda}}}{\partial\lambda_i}=\frac{1}{2}(\rho_{\boldsymbol{\lambda}}\mathcal{L}_i+\mathcal{L}_i\rho_{\boldsymbol{\lambda}}) \,,
\end{equation}
so that the matrix elements of the QFIm are
\begin{equation}\label{QFISLD}
  [\mathcal{K}(\boldsymbol{\lambda})]_{ij}
  = \frac{1}{2}\,\text{Tr} \left[
      \rho_{\boldsymbol{\lambda}}
      \{\mathcal{L}_i,\mathcal{L}_j\}
    \right],
\end{equation}
with $\{\cdot,\cdot\}$ the anticommutator. 
To understand (\ref{QFISLD}), recall that, given a manifold of density operators such as $\mathcal{S}$, one can quantify the distinguishability between any two such operators by means of the Bures distance \cite{BuresOG,HUBNER1992239},
\begin{equation}\label{Bures1}
\begin{aligned}
    d_B^2(\rho_1,\rho_2)&=2(1-\text{Tr}[(\sqrt{\rho_1}\rho_2\sqrt{\rho_1})^{1/2}])\\
    &=2\left(1-\sqrt{F(\rho_1,\rho_2)}\right)\,,
\end{aligned}
\end{equation}
where $F(\rho_1,\rho_2)$ denotes the fidelity between the two states $\rho_1$ and $\rho_2$ \cite{UHLMANN1976273,Allardyce:87}. 
It was shown by Braunstein and Caves \cite{PhysRevLett.72.3439} (building on the work of H\"ubner \cite{HUBNER1992239})  that the Bures distance between
the nearby states $\rho_{\boldsymbol{\lambda}}$ and
$\rho_{\boldsymbol{\lambda}+d\boldsymbol{\lambda}}$ admits the expansion
\begin{equation}\label{Bures2}
\begin{aligned}
    d_B^2 \left(
        \rho_{\boldsymbol{\lambda}},
        \rho_{\boldsymbol{\lambda}+d\boldsymbol{\lambda}}
    \right)
    &= \frac{1}{4}
      \sum_{i,j}
      \text{Tr} \left[
        \rho_{\boldsymbol{\lambda}}
        \{\mathcal{L}_i, \mathcal{L}_j\}
      \right]
      d\lambda_i d\lambda_j \\
    &= \frac{1}{4}\,
      d\boldsymbol{\lambda}^{T}
      \mathcal{K}(\boldsymbol{\lambda})
      d\boldsymbol{\lambda} 
    = \frac{1}{4} ds_\rho^{2},
\end{aligned}
\end{equation}
where $\mathcal{K}(\boldsymbol{\lambda})$ is the QFIm
 defined by Eq.~(\ref{QFISLD}).

We now restrict our attention to a special class of quantum models, which are obtained by applying a quantum channel depending on the unknown parameter $\boldsymbol{\lambda}$ to a given initial probe state. 
Let $\Gamma_{\boldsymbol{\lambda}}$ be a quantum channel (completely
positive trace-preserving map) with Kraus representation
$\{K_j(\boldsymbol{\lambda})\}_{j=1}^N$ such that
\begin{equation}
\label{KraussRepresentation}
    \rho_{\boldsymbol{\lambda}}
    = \Gamma_{\boldsymbol{\lambda}}(\rho_0)
    = \sum_{j=1}^{N}
      K_j(\boldsymbol{\lambda})\,\rho_0\,
      K_j^{\dagger}(\boldsymbol{\lambda}).
\end{equation}
For an infinitesimal parameter shift
$\boldsymbol{\lambda}\to\boldsymbol{\lambda}+d\boldsymbol{\lambda}$, 
one can define a {\em metrology matrix} \cite{PhysRevA.96.012310} as the matrix $M(\boldsymbol{\lambda},d\boldsymbol{\lambda})$ with elements
\begin{equation}\label{metrologymatrixelements}[M(\boldsymbol{\lambda},d\boldsymbol{\lambda})]_{ij}=\text{Tr}[K_i(\boldsymbol{\lambda})^\dagger K_j(\boldsymbol{\lambda}+d\boldsymbol{\lambda})\rho_0].
\end{equation}
It then follows \cite{PhysRevA.96.012310} 
that the fidelity between the nearby
states $\rho_{\boldsymbol{\lambda}}$ and
$\rho_{\boldsymbol{\lambda}+d\boldsymbol{\lambda}}$ can be written compactly as the trace norm of this matrix (at leading order with respect to $d\boldsymbol{\lambda}$) ,
\begin{eqnarray}
    F(\rho_{\boldsymbol{\lambda}},\rho_{\boldsymbol{\lambda}+d\boldsymbol{\lambda}})=\left \lVert M(\boldsymbol{\lambda},d\boldsymbol{\lambda}) \right \rVert_1 + o(d\boldsymbol{\lambda}^2).
\end{eqnarray}
Comparing this with Eqs.~(\ref{Bures1}), (\ref{Bures2}), and using
\begin{eqnarray}
    F(\rho_{\boldsymbol{\lambda}},\rho_{\boldsymbol{\lambda}+d\boldsymbol{\lambda}})= 1 - \tfrac18\,d\boldsymbol{\lambda}^{T}\mathcal{K}(\boldsymbol{\lambda})d\boldsymbol{\lambda},
\end{eqnarray}
(at leading order), one obtains an elegant representation of the QFIm:
\begin{equation}
\label{QFImetrologymatrix}
    d\boldsymbol{\lambda}^{T}\,
    \mathcal{K}(\boldsymbol{\lambda})\,
    d\boldsymbol{\lambda}
    = 8\bigl(1 - \| M (\boldsymbol{\lambda},d\boldsymbol{\lambda})\|_{1}\bigr) + o(d\boldsymbol{\lambda}^2).
\end{equation}
This form is particularly useful when the parameter dependence enters through a known
dynamical evolution or transfer function, as it allows the QFIm to be computed
directly from the derivatives of the Kraus operators without explicitly
evaluating the SLDs.

\subsubsection{Classical and quantum semi-parametric estimation}\label{SubSectionSemiP}

In certain metrological scenarios, one is interested in precisely estimating only a small
subset of all the parameters encoded in a probe state.  Let the full parameter
vector be partitioned as $\boldsymbol{\lambda}=(\beta,\boldsymbol{\eta})$, where the scalar
$\beta$ is the parameter of interest and the remaining components
$\boldsymbol{\eta}$ play the role of nuisance parameters.  
If the full (classical or quantum) Fisher information matrix is 
invertible, then we can bound the error on estimating the parameter of interest $\beta$ by means of a suitable reparametrization of the CRB (\ref{QCRB}) \cite{doi:10.1142/S0219749909004839}: 
\begin{equation}
\label{QCRBB}
\begin{aligned}
    \text{Var}_{\boldsymbol{\lambda}}[\hat{\beta}]&\geq (\boldsymbol{\partial\beta})^T\mathcal{I}(\boldsymbol{\lambda};\{\Pi_\omega\})^{-1}\boldsymbol{\partial\beta} \\
    &\geq(\boldsymbol{\partial\beta})^T\mathcal{K}(\boldsymbol{\lambda})^{-1}\boldsymbol{\partial\beta}\equiv\mathcal{V}^{\rm Q}_\beta,
    \end{aligned}
\end{equation}
where $\hat{\beta}$  is an unbiased estimator and $\boldsymbol{\partial\beta}$ is a vector of derivatives $[\boldsymbol{\partial\beta}]_i=\partial\beta/\partial\lambda_i$  with respect to the parameter set $\boldsymbol{\lambda}$. 
However, when the original parameter set $\boldsymbol{\lambda}$ is high-dimensional or even
infinite-dimensional, the method in (\ref{QCRBB}) becomes
intractable.  Invoking the theory of {\it semi-parametric estimation} offers a powerful way forward.

Consider the scenario where the measurement 
$\{\Pi_\omega\}$ is performed repeatedly on $m$ quantum samples 
$\rho_{\boldsymbol{\lambda}}$ resulting in independent outcomes 
$\omega_1,\dots, \omega_m$  drawn from $p_{\boldsymbol{\lambda}}(\omega)$. Let $\boldsymbol{\lambda}_0=(\beta_0, \boldsymbol{\eta}_0 )$ be a fixed parameter value and assume that the estimator  $\hat{\beta}_m = \hat{\beta}_m(\omega_1,\dots \omega_m)$ is (locally) unbiased around $\boldsymbol{\lambda}_0$ and can be approximated as an empirical average 
\begin{equation}\label{RAL}
    \hat{\beta}_m - \beta_0
    \approx   \frac{1}{m} \sum_{i=1}^m \varphi(\omega_i)
\end{equation}
for some function $\varphi$ called the \emph{influence function} (IF) \cite{Tsiatis}.  
Intuitively, the IF quantifies how much each measurement outcome ``pushes'' the
estimator away from the reference value $\beta_0$. The local unbiasedness requirement translates 
into the first order expansion
\begin{equation}
\label{eq:local.unbiased.phi}
\mathbb{E}_{\boldsymbol{\lambda}}[\varphi] = \beta -\beta_0 + o(\beta -\beta_0).
\end{equation}
and in particular, $\mathbb{E}_{\boldsymbol{\lambda}_0}[\varphi] =0 $.
By the central limit theorem at $\boldsymbol{\lambda}_0$, the LHS of Eq.~(\ref{RAL})  
converges in distribution to the centered normal distribution
\begin{equation}
    \sqrt{m}\, (\hat{\beta}_m - \beta_0)
    \xrightarrow{d}
    \mathcal{N}(0,\,\|\varphi\|_{\boldsymbol{\lambda}_0}^2),
\end{equation}
where $\|\varphi\|_{\boldsymbol{\lambda}_0}^2 = \mathbb{E}_{\boldsymbol{\lambda}_0}[\varphi^2]$ 
so that
\begin{equation}\label{HB0}
 \lim_{m\to \infty}  m \mathrm{Var}_{\boldsymbol{\lambda}_0}[\hat{\beta}_m]
   = \|\varphi\|_{\boldsymbol{\lambda}_0}^2.
\end{equation}
Thus, the asymptotic error of any estimator of the form \eqref{RAL} is determined entirely by
the norm of its IF.

This viewpoint leads naturally to a geometric picture.  
Let 
\begin{equation}
\label{eq:Hilbert.space.H}
\mathcal{H}_{\boldsymbol{\lambda}_0} := \{ g (\omega): \mathbb{E}_{\boldsymbol{\lambda}_0}[g] =0, \mathbb{E}_{\boldsymbol{\lambda}_0}[g^2]<\infty\}
\end{equation}
denote the space of zero-mean,
square-integrable (complex) functions of the data, and equip it with the Hilbert space structure defined by the inner product
$\langle g_1,g_2\rangle_{\boldsymbol{\lambda}_0} = \mathbb{E}_{\boldsymbol{\lambda}_0} [\overline{g_1}g_2]$.  
The norm in this space is the variance, $\|g\|_{\boldsymbol{\lambda}_0}^2=\text{Var}_{\boldsymbol{\lambda}_0}[g]$.  
The score functions
\begin{equation} \label{thescorefunctions}
 S_{\boldsymbol{\lambda}_0,i}(\omega):=\left.\frac{\partial[\log p_{\boldsymbol{\lambda}}(\omega)]}{\partial{\lambda_i}}
\right|_{\boldsymbol{\lambda}_0}
\end{equation}
span the co-tangent space
\begin{equation}\label{tangentspace}
    \mathcal{T}_{\boldsymbol{\lambda}_0}
    \,= \text{span}\{S_1,\dots,S_n\}\subseteq\mathcal{H}_{\boldsymbol{\lambda}_0}.
\end{equation}
From Eq.~\eqref{eq:local.unbiased.phi} we find that the set of 
IFs for estimating the target parameter $\beta$ is
\begin{equation}
\label{GeneralIFrelation}
    \Phi_{\boldsymbol{\lambda}_0}
    = \{\varphi\in\mathcal{H}_{\boldsymbol{\lambda}_0} :
       \langle \varphi, S_i\rangle_{\boldsymbol{\lambda}_0}
       = \partial\beta/\partial\lambda_i, \quad
       \forall i=1,\ldots,n \}.
\end{equation}
According to Eq.~\eqref{HB0}, for each such IF the rescaled asymptotic mean square error of the corresponding estimator 
$\hat{\beta}_m$ is $\|\varphi\|_{\boldsymbol{\lambda}_0}^2$.
The smallest possible error is achieved by the \emph{efficient influence function} $ \varphi_{\rm eff}$ given by the unique orthogonal
projection of any $\varphi\in\Phi$ onto the co-tangent space $\mathcal{T}_{\boldsymbol{\lambda}_0}$,
\begin{equation}
\label{eq:projection.tangent.space}
    \varphi_{\rm eff}=\mathrm{Proj}_{\mathcal{T}_{\boldsymbol{\lambda}_0}
    }(\varphi),
\end{equation}
such that 
\begin{equation}\label{GHB}
    \left \lVert \varphi_{\rm eff} \right \rVert_{\boldsymbol{\lambda}_0}^2 =
    \min_{\varphi\in\Phi}\left \lVert \varphi \right \rVert_{\boldsymbol{\lambda}_0}^2.
\end{equation}
By setting $\varphi_{\rm eff}=\sum_j c_j S_j$ and solving the constraints (\ref{GeneralIFrelation}), we get
$\boldsymbol{c}_{\rm eff}=\mathcal{I}(\boldsymbol{\lambda};\{\Pi_{\omega}\})^{-1}\boldsymbol{\partial\beta}$, and find that the corresponding estimator attains the classical Cram\'er-Rao bound --- first inequality in \eqref{QCRBB} --- in an asymptotic sense,
\begin{equation}
\label{GHBB}
\lim_{m\to \infty} m {\rm Var}_{\boldsymbol{\lambda}_0} [\hat{\beta}_m]
    =
    \|\varphi_{\rm eff}\|_{\boldsymbol{\lambda}_0}^2
    = 
   (\boldsymbol{\partial\beta})^{T}
      \mathcal{I}(\boldsymbol{\lambda};\{\Pi_\omega\})^{-1}  \boldsymbol{\partial\beta}.
\end{equation}

Moving to a more general non-parametric setting where the parameter $\boldsymbol{\lambda}$ is not finite-dimensional, the approach via the Fisher information is problematic, while 
the IF analysis can still provide the optimal estimation error for one-dimensional (and more generally finite-dimensional) parameters $\beta$. 
The key point is the identification of the co-tangent space $\mathcal{T}_{\boldsymbol{\lambda}}$ as the closure in 
$\mathcal{H}_{\boldsymbol{\lambda}}$ of the span of score functions obtained by constructing ``smooth'' one-parameter sub-models passing through $\boldsymbol{\lambda}$. Influence functions can be defined by generalizing the condition \eqref{eq:local.unbiased.phi} or \eqref{GeneralIFrelation} appropriately, and have a unique efficient IF projection onto the co-tangent space. The leftmost equality in \eqref{GHBB} holds for the efficient IF even if the Fisher information interpretation is not available. For more details on semi-parametric estimation for infinite-dimensional models we refer to \cite{Tsiatis,Bickel,PhysRevX.10.031023,PhysRevResearch.1.033006}.


Until now the measurement has been fixed and we 
analyzed an essentially classical estimation problem for the model $p_{\boldsymbol{\lambda}}(\omega) = {\rm Tr}(\rho_{\boldsymbol{\lambda}}\Pi_\omega)$. Fortunately, the construction can be lifted to the fully quantum level \cite{PhysRevX.10.031023}, allowing us to identify optimal measurement and estimation procedures which achieve 
the bound $\mathcal{V}^{\rm Q}_\beta$ in \eqref{QCRBB}. Briefly, we consider the (real) Hilbert space 
\begin{equation}
\mathcal{H}^Q_{\boldsymbol{\lambda}_0}
:=\{A : {\rm Tr}(\rho_{\boldsymbol{\lambda}_0} A) =0, {\rm Tr}(\rho_{\boldsymbol{\lambda}_0}A^2) <\infty\}  
\end{equation}
with inner product 
$\langle A, B\rangle_{\boldsymbol{\lambda_0}} := \frac{1}{2}{\rm Tr}( \rho_{\boldsymbol{\lambda_0}}
\{A, B\})
$
where $A,B$ are (possibly unbounded) selfadjoint operators on $\mathcal{H}$. 
The co-tangent space is  
\begin{equation}
\mathcal{T}^Q_{\boldsymbol{\lambda}_0}
:=
\text{span}
\{\mathcal{L}_1,\dots, \mathcal{L}_n\} \subset \mathcal{H}^Q_{\boldsymbol{\lambda}_0}
\end{equation}
where $\mathcal{L}_i$ are the SLD (quantum score) operators defined in 
Eq.~\eqref{SLD}. An influence operator is an element  of 
\begin{equation}
\Phi^Q_{\boldsymbol{\lambda}_0}:=
\{
H\in \mathcal{H}^Q_{\boldsymbol{\lambda}_0} : 
\langle H, \mathcal{L}_i \rangle_{\boldsymbol{\lambda}_0 }= \partial \beta/\partial \lambda_i,\quad \forall i=1, \dots, n\}.
\end{equation}
By repeatedly measuring an influence operator $H$ on $m$ copies of 
$\rho_{\boldsymbol{\lambda}}$, we can construct an average estimator $\hat{\beta}_m$ as in \eqref{RAL} whose rescaled asymptotic error is 
$\|H\|_{\boldsymbol{\lambda}_0}^2$ (for $\boldsymbol{\lambda}$ close to $\boldsymbol{\lambda}_0$). As in the classical case, the smallest error is achieved by the efficient influence operator $H_{\rm eff}$ which is the unique projection of any influence operator onto $\mathcal{T}^Q_{\boldsymbol{\lambda}_0}$
\begin{equation}
H_{\rm eff}:= {\rm Proj}_{\mathcal{T}^Q_{\boldsymbol{\lambda}_0}} (H)
\end{equation}
This can be computed explicitly as $H_{\rm eff} = \sum h_i \mathcal{L}_i$ with $\boldsymbol{h}:= \mathcal{K}(\boldsymbol{\lambda}_0)^{-1}\boldsymbol{\partial \beta}$ and then we obtain 
\begin{equation}\label{QCRBSEMI}
\lim_{m\to \infty} m {\rm Var}_{\boldsymbol{\lambda}_0} (\hat{\beta}_m)
    =
    \|H_{\rm eff}\|_{\boldsymbol{\lambda}_0}^2
     = \boldsymbol{\partial\beta}^T\mathcal{K}(\boldsymbol{\lambda})^{-1}
      \boldsymbol{\partial\beta},
\end{equation}
which shows the achievability of the scalar-parameter bound (\ref{QCRBB}) and provides the semi-parametric framework for quantum estimation with 
arbitrarily many nuisance parameters. 

\subsection{Imaging theory}\label{SectionImagingTheory}

\begin{figure}[htp]
\raggedright
\begin{subfigure}{0.45\textwidth}
    \includegraphics[width=\linewidth]{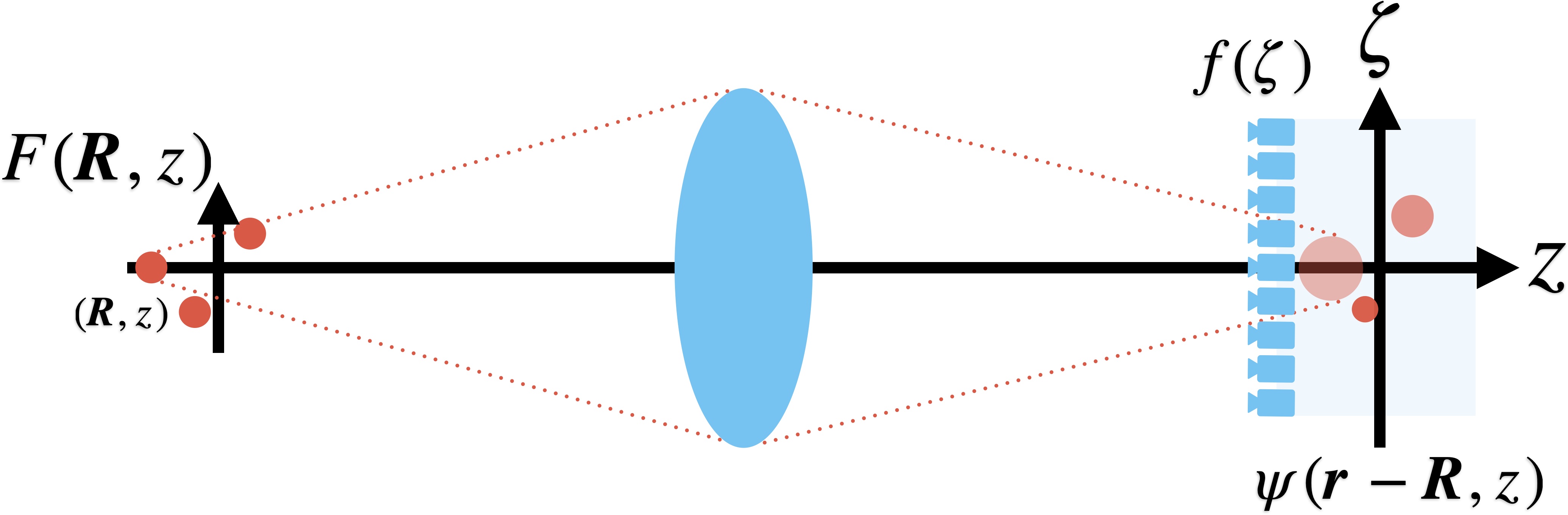}
    \caption{}
    \label{FigureA}
\end{subfigure}

\begin{subfigure}{0.45\textwidth}
    \includegraphics[width=\linewidth]{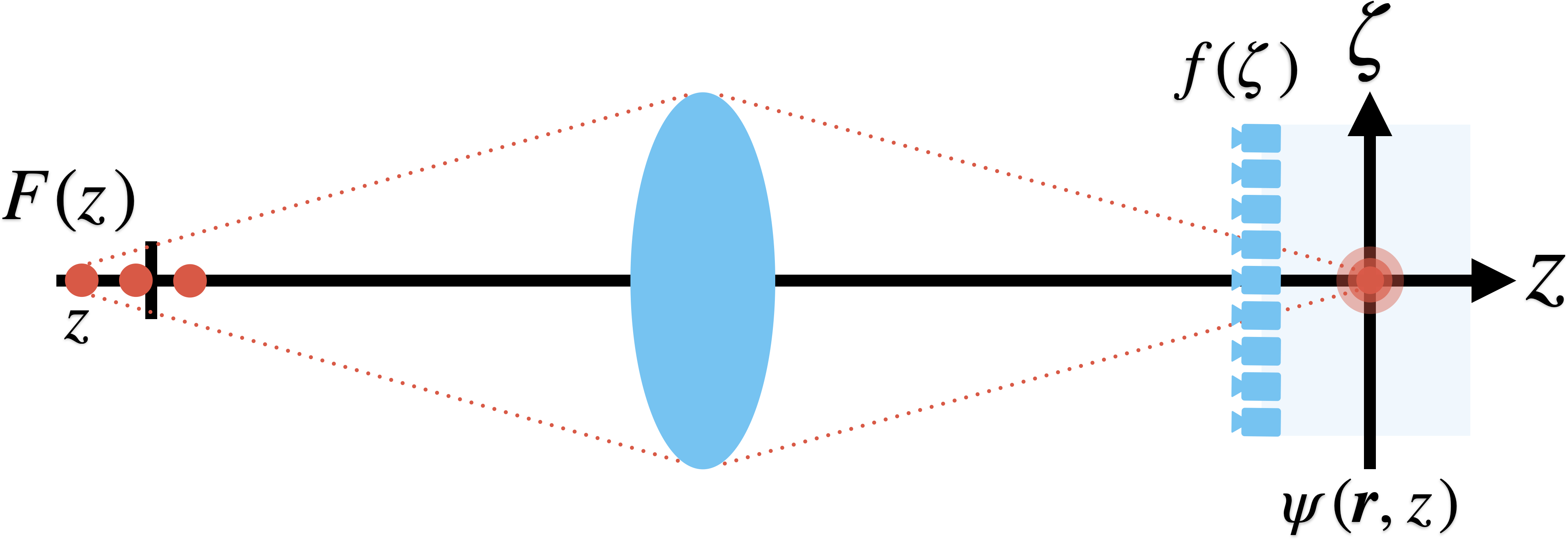}
    \caption{}
    \label{FigureB}
\end{subfigure}
\caption{{Panel (a)}: A schematic of the full imaging system consisting of a rough surface, modelled as $N$ discrete light sources displaced axially and radially from the focal point of the lens characterized by a point spread function $\psi$. The incoming field is then measured using an appropriate measurement scheme (represented as the field of cameras in the image) yielding data in the representation $\zeta$ with intensity $f(\zeta)$. {Panel (b)}: A special case of the model in {(a)}, consisting of sources only axially displaced from focus.\label{Figure2}}
\end{figure}

The general model we consider in this paper  is illustrated in Figure~\ref{FigureA} and consists of single photon emitters at positions $(x,y,z)\equiv(\boldsymbol{R},z)$ near the object plane, distributed according to an unknown probability distribution $F$, and a lens that focuses the point sources onto the image plane with radial coordinate \red{$\boldsymbol{r}=(u, \varv)$}, where a measurement takes place. 

A central aspect of this paper involves understanding the optimal information that we can extract through different measurement strategies. To this end, let us describe our imaging model in more detail. Like much of the literature \cite{tsang2019resolving}, we assume weak sources. That is, the state of the focused light holds significant contributions only from the zero photon state and the single photon state. With this approximation, we are able to treat a photon in the image plane that originates from the focus of the imaging system as the pure state
\begin{eqnarray}
\label{FocusState}
    \ket{\psi_0}=\int\psi(\boldsymbol{r})\ket{\boldsymbol{r}}d\boldsymbol{r}.
\end{eqnarray}
Here $\psi(\boldsymbol{r})$ is the \textit{point spread function} (PSF) of the imaging system and describes how the field emitted from a single point source in focus appears after passing through the lens. One can see from (\ref{FocusState}) that the PSF gives a weighting to the continuous distribution of the states representing the creation of a single photon at position $\boldsymbol{r}$ in the image plane, $\ket{\boldsymbol{r}}$. If an emitter is displaced from focus then this will have effects on their images. The state of a photon at the image plane originating from an emitter displaced by coordinates $(\boldsymbol{R},z)$ can be written as
\begin{equation}
\begin{aligned}
    \ket{\psi_{\boldsymbol{R},z}}&=\red{e^{-(x\partial_{u}+y\partial_{\varv}+i zG)}}\ket{\psi_0}\\
    &=\int\psi(\boldsymbol{r}-\boldsymbol{R},z)\ket{\boldsymbol{r}}d\boldsymbol{r},
    \end{aligned}
\end{equation}
where
\begin{align}\label{introgenerators}
    \psi(\boldsymbol{r}-{\boldsymbol{R},z})&\equiv \red{e^{-(x\partial_{u}+y\partial_{\varv}+i zG)}}\psi(\boldsymbol{r}).
\end{align}
The generators \red{$P_x=-i\partial_u$ and $P_y=-i\partial_\varv$} account for the radial displacements of the source $\boldsymbol{R}=(x,y)$ and radially displace the PSF in the image plane without deforming it. The generator $G$ however, induces broadening as well as a complex phase profile in the PSF, due to the source being ``out of focus" by an amount $z$. Implicitly, $G$ is  defined via the paraxial wave equation --- to which our PSF is a solution, 
\begin{equation}
\label{ParaxialWaveEquation}
   \frac{\partial\psi}{\partial z} =-i G\psi\equiv -i\frac{1}{2k}\nabla^2_{\perp}\psi,
\end{equation}
where \red{$\nabla^2_{\perp}=\partial^2/\partial u^2 + \partial^2/\partial \varv^2$} represents the Laplacian only in the transverse plane and $k$ represents the wave number of the field. In this work we will consider distributions of sources $F(\boldsymbol{R},z)$ and thus the state of a photon at the image plane can be written as the mixture,
\begin{equation}
\label{statistical mixture}
    \rho=\int\ket{\psi_{\boldsymbol{R},z}}\bra{\psi_{\boldsymbol{R},z}}dF(\boldsymbol{R},z).
\end{equation}
The PSFs we use to exemplify our results throughout this work amount to Gaussian beams. We content ourselves with such a representative choice due to the similarities between the Gaussian beam, originating from a ``Gaussian aperture'', and the Airy beam which is the exact solution to the circular aperture. The general form of a normalized Gaussian beam at the image plane is
\begin{eqnarray}
\label{Image plane PSF}
    \psi(\boldsymbol{r}-\boldsymbol{R},z)=\sqrt{\frac{2}{\pi}}\frac{1}{\omega(z)}e^{-\frac{|\boldsymbol{r}-\boldsymbol{R}|^2}{\omega(z)^2}}e^{\frac{-ik|\boldsymbol{r}-\boldsymbol{R}|^2}{2R(z)}}.
\end{eqnarray}
Here we have defined
\begin{eqnarray}
    \frac{1}{R(z)}=\frac{z}{z^2+z_R^2},
\end{eqnarray}
where the Rayleigh range $z_R=k\omega_0^2/2$ is a scaling parameter that describes how the beam width, defined as
\begin{eqnarray}
    \omega(z)=\omega_0\sqrt{1+\frac{z^2}{z_R^2}},
\end{eqnarray}
changes with distance from its value $\omega_0$ at focus. In what follows we set $\omega_0=1$ with no loss of generality. 

A significant portion of this paper is devoted to comparing the optimality of two different manners of measuring the incoming light field. The projectors representing these schemes are denoted by $\{\Pi_\zeta\} \equiv \{\ket{\zeta}\bra{\zeta}\}$ where $\zeta$ denotes the measure representing the result of a particular measurement. As described in section~\ref{SectionMetrologyTheory}, the probability of obtaining $\zeta$ by measuring the system in the state $\rho$ in the $\ket{\zeta}$ basis is given by
\begin{equation}
    \label{f}
    f(\zeta)=\text{Tr}\left[\rho\ket{\zeta}\bra{\zeta}\right]
    =\int H(\zeta;\boldsymbol{R},z) dF(\boldsymbol{R},z),
\end{equation}
where $H(\zeta;\boldsymbol{R},z)$ represents the probability of measuring the field $\ket{\psi}$ in the basis spanned by $\ket{\zeta}$,
\begin{equation}
\label{H}
    H(\zeta;\boldsymbol{R},z)=|\sprod{\zeta}{\psi_{\boldsymbol{R},z}}|^2.
\end{equation}

The two measurement schemes we compare in this work are \textit{direct imaging}  and \textit{spatial mode demultiplexing} (SPADE). Direct imaging is an intensity measurement on the image plane, and can be implemented in practice via pixel-by-pixel photon counting, much like a standard camera would do. Here $\zeta\equiv\boldsymbol{r}$ corresponds to a position on the photosensitive screen (pixel basis). On the other hand, a SPADE setup \cite{PhysRevX.6.031033} decomposes the incident light field into a set of orthonormal modes before measuring the intensity in each mode. The modes' wavefunctions must have known parity and be real valued \cite{Rehacek:17}. In SPADE, the measurement basis label $\zeta\equiv q$ indicates which transverse electromagnetic mode we are measuring \cite{Lasers}. SPADE can be thought of intuitively as ``Taylor series in action''. When a point source is displaced {\it radially} from focus, its PSF displaces also. Thus one can extract more information by measuring in a basis which also captures this displacement. For example, if we consider a radially displaced Gaussian PSF, we find that the first few terms in a Taylor expansion of the PSF correspond to the Hermite-Gauss basis \cite{PhysRevX.6.031033}. For {\it axial} displacements,  the intuition is slightly less obvious. In this case, the PSF doesn't displace but rather distorts itself based on how out of focus the point source is.

\red{As a practically relevant special case of the general three-dimensional framework, for the majority of this paper we will focus on the reduced problem} of imaging a tiny surface patch, akin to a ``rabbit hole'', see  Figure~\ref{FigureB}. In this model we consider {\it only} sources displaced axially --- that is towards or away from the observer along the optical axis but not in the plane tangential to the lens. 
At the image plane, the Gaussian beam thus reduces to
\begin{equation}
    \label{Reduced Image plane PSF}
    \psi(\boldsymbol{r},z)=\sqrt{\frac{2}{\pi}}\frac{1}{\omega(z)}e^{\frac{-r^2}{\omega(z)^2}}e^{\frac{-ikr^2}{2R(z)}}.
\end{equation}
and Eqs.~(\ref{f}) and (\ref{H}) reduce to
\begin{equation}
    \label{Reduced f}
    f(\zeta)=\int H(\zeta;z) dF(z),
\end{equation}
and 
\begin{equation}
\label{Reduced H}
    H(\zeta;z)=|\sprod{\zeta}{\psi_{z}}|^2.
\end{equation}

\section{Ultimate limits: Computing \texorpdfstring{\\}{} the Quantum Fisher Information}\label{SectionComputingQFI}
In this section we compute the ultimate precision limits on estimating \red{moments of the three-dimensional distribution of an arbitrary discrete number $N$ of weak incoherent point sources.} This includes in particular a tight upper bound on the information achievable in roughness estimation for the \red{general surface metrology problem described in section~\ref{SectionImagingTheory} and illustrated in Figure~\ref{FigureA}. This result is then specialized to the reduced problem of estimating the first and second moments of a purely axial distribution as illustrated in Figure~\ref{FigureB}}. 

\red{
The state of the system (\ref{statistical mixture}) as a function of the source displacements $\boldsymbol{D}=\left(\boldsymbol{D}_1,\boldsymbol{D}_2,\ldots,\boldsymbol{D}_N\right)^T$, where $\boldsymbol{D}_i=(\boldsymbol{R}_i,z_i)=(x_i,y_i,z_i)$, can be described using the Kraus representation for the channel corresponding to the PSF propagation,
\begin{equation}
    \rho_{\boldsymbol{D}}=\sum_j {K}_j(\boldsymbol{D})\rho_{0} K_j(\boldsymbol{D})^{\dagger},
\end{equation}
where $\rho_0:= \ket{\psi_0}\bra{\psi_0}$ with $\ket{\psi_0}$ defined  in (\ref{FocusState}), and the Kraus operators are of the form
\begin{equation}
    K_j(\boldsymbol{D})=\sqrt{p_j}e^{-i\boldsymbol{D}_j\cdot \boldsymbol{\mathfrak{G}}},
\end{equation}
with $p_j$ being the relative intensity of the $j^{\text{th}}$ source, and $\boldsymbol{\mathfrak{G}}=(P_x,P_y,G)$ denoting the vector of generators describing the state propagation from source to image plane as in  Eq.~(\ref{introgenerators}).}

The metrology matrix \eqref{metrologymatrixelements} is 
\red{
\begin{eqnarray}
  [M(\boldsymbol{D}, \boldsymbol{dD})]_{ij} 
  &=&
  {\rm Tr}\left[ K_i(\boldsymbol{D})^\dagger
   K_j(\boldsymbol{D}+ \boldsymbol{dD}) \rho_0\right]
   \nonumber
   \\
   &=&
   \langle K_i(\boldsymbol{D})\psi_0| K_j(\boldsymbol{D}+ \boldsymbol{dD})\psi_0\rangle.
\end{eqnarray}}
As we are interested only in displacements in the sub-diffraction regime, we assume \red{
$|\boldsymbol{D}_j|\ll 1$ and thus in the limit 
$\boldsymbol{D}\to \boldsymbol{0}$ we have
\begin{equation}
\begin{aligned}
\lim_{\boldsymbol{D}
\to \boldsymbol{0}} \,
[M(\boldsymbol{D}, \boldsymbol{dD})]_{ij} &= 
[M(\boldsymbol{0}, \boldsymbol{dD})]_{ij}  \\ &=
\sqrt{p_ip_j}\langle \psi_0| e^{-i \boldsymbol{dD}_j \cdot \boldsymbol{\mathfrak{G}}}\psi_0\rangle,
\end{aligned}
\end{equation}
so that 
\begin{equation}
[M(\boldsymbol{0}, \boldsymbol{dD})^\dagger M(\boldsymbol{0}, \boldsymbol{dD}) ]_{ij} = \bar{c}_i c_j,
\end{equation}
where $c_j := \sqrt{p}_j \langle\psi_0 | e^{-i \boldsymbol{dD}_j\cdot\boldsymbol{\mathfrak{G}}}\psi_0\rangle$. Therefore 
\begin{eqnarray}
\|M(\boldsymbol{0}, \boldsymbol{dD})\|_{1} &=&
\|\boldsymbol{c}\| = \sqrt{\sum_j |c_j|^2} 
\nonumber
\\
&=&
\left(\sum_{j=1}^N p_j\left(1-\text{Cov}_{{0}}[\boldsymbol{dD}_j\cdot\boldsymbol{\mathfrak{G}}]+o(\boldsymbol{dD}^2)
\right)\right)^{1/2} \nonumber\\
&=&
1 -\frac{1}{2}\sum_j p_j \text{Cov}_{{0}}[\boldsymbol{dD}_j\cdot\boldsymbol{\mathfrak{G}}].
\label{MetMatrixTraceNorm}
\end{eqnarray}}

Using Eq.~(\ref{QFImetrologymatrix}), we find that for very small distributions around the origin \red{($\boldsymbol{D}\approx \boldsymbol{0}$)
the QFIm is, at leading order, a $3N\times 3N$ matrix
\begin{equation}
\label{eq:QFIdisplacements}
    \mathcal{K}(\boldsymbol{D}\approx\boldsymbol{0})=4\ {\rm{diag}}(p_1, \dots, p_N)\otimes\text{Cov}_{{0}}[\boldsymbol{\mathfrak{G}}],
\end{equation}
where $\text{Cov}_{{0}}[\boldsymbol{\mathfrak{G}}]$ is the $3\times3$ covariance matrix of the vectorial generator $\boldsymbol{\mathfrak{G}}$. We note that, since expectations are taken with respect to $\ket{\psi_0}$ and, as $\psi(\boldsymbol{r})$ is real and symmetric about the focal point of the imaging system, we have $\langle P_x\rangle_0=\langle P_y\rangle_0=0$ and also $\langle P_x G\rangle_0=\langle P_y G\rangle_0=\langle P_x P_y\rangle_0=0$. Thus the off-diagonal elements of $\text{Cov}_0[\boldsymbol{\mathfrak{G}}]$ vanish and, in particular, for sources of equal intensity the QFIm is diagonal, 
\begin{equation}\label{3dqfim}
\mathcal{K}(\boldsymbol{D}\approx\boldsymbol{0})=4 \frac{\mathbb{I}_N}{N}\otimes{\rm{diag}}\big(\Delta_0P_x^2,\Delta_0P_y^2,\Delta_0G^2\big),
\end{equation} where $\Delta_0A^2:=\langle A^2\rangle_0-\langle A\rangle_0^2$.}

\red{Equation (\ref{3dqfim}) shows that, in the sub-diffraction regime, the three spatial directions decouple at leading order for symmetric imaging systems, with the attainable precision in each direction determined solely by the variance of the corresponding generator.} 
While this result provides important insight into \red{the three-dimensional localization problem, the use of source coordinates $\boldsymbol{D}$ as estimation parameters} makes it hard to extrapolate to more general models involving continuous source distributions. In addition, our interest is in estimating the surface roughness which is naturally related to a characterization in terms of \red{spatial} moments. For these reasons we will recast the above result in terms of this alternative parametrization. 
\red{Let $\boldsymbol{\theta} :=\{\theta_{l_\alpha}\}_{l=1}^N$ be the moments given by
\begin{eqnarray}
\label{DiscreteMoments}
    \theta_{l_\alpha}=\sum_{k=1}^N p_k D_{k_\alpha}^l,
\end{eqnarray}
where we have introduced the index $\alpha\in\{x,y,z\}$ to consider --- in more generality than the rest of the paper --- moments of both  axial ($\alpha=z$) \textit{and} radial displacements ($\alpha \in \{x,y\}$) of the source distribution.}
Then the inverse QFIm in the 
$\boldsymbol{\theta}$ parametrization is
\red{\begin{eqnarray}
\label{QFIMomentError}
    \mathcal{K}(\boldsymbol{\theta})^{-1}=
    J^{T} 
    \mathcal{K}(\boldsymbol{D})^{-1}
    J
\end{eqnarray}}
where $J$ is the Jacobian
\begin{equation}
 \red{   J_{i_\alpha ,j_\beta} = \frac{\partial\theta_{j_\beta}}{\partial D_{i_\alpha}} = j p_i D_{i_\alpha}^{j-1}\delta_{\alpha\beta},}
\end{equation}
with $\theta_{0_\alpha}:=1$ \red{and $\delta_{\alpha\beta}$ the Kronecker delta}. By  inverting \eqref{eq:QFIdisplacements} 
we find that, for 
\red{$|\boldsymbol{D}|\ll 1$, 
\begin{eqnarray}\label{eq:inverseQFI.theta}
    [\mathcal{K}(\boldsymbol{\theta})^{-1}]_{i_\alpha,j_\beta} \approx \frac{ij\delta_{\alpha\beta}}{4\, \text{Cov}_0[\mathfrak{G}]_{\alpha\beta}}\theta_{(i+j-2)_\alpha}.
\end{eqnarray}}

According to \eqref{QCRBSEMI}, the diagonal elements of \eqref{eq:inverseQFI.theta} are the asymptotic errors of estimating the corresponding moments.  
\red{Note that  
$\theta_{j_\alpha} = \mathcal{O}(\Delta^j
)$, where $\Delta$ is the ``object size'' \cite{PhysRevA.99.013808} 
\begin{equation}\label{objectsize}
\Delta \propto \max_{j \neq k} \mbox{$\sqrt{\sum_\alpha |\boldsymbol{D}_{j_\alpha}-\boldsymbol{D}_{k_\alpha}|^2}$},
\end{equation}
which can also be referred to as the  ``feature depth'' in the purely axial reduced case, $\Delta \propto \max_{j \neq k} |z_j-z_k|$. 
This means that the 
moments of order $l\geq 2$ along each individual direction $\alpha$ can be estimated with errors 
$\mathcal{V}^{\rm Q}_{\theta_{l_\alpha}}= \mathcal{O}(\Delta^{2l-2})$, while the mean of the axial distribution $\mu \equiv \theta_{1_z}$ has error} 
\begin{equation}
\label{M1QFI}
\mathcal{V}^{\rm Q}_{\mu}= \frac{1}{4\Delta_0 G^2},
\end{equation}
which does not vanish in the sub-diffraction limit $\Delta\to 0$. 

\red{We observe that all the matrix elements corresponding to different directions $\alpha \neq \beta$ in Eq.~(\ref{eq:inverseQFI.theta}) 
vanish. We thus conclude that, in the sub-diffraction limit $\boldsymbol{D}\rightarrow\boldsymbol{0}$, the individual moments characterizing source distributions along each of the three coordinate axes of a three-dimensional point source cloud are statistically independent. This means we can easily extract the fundamental error bounds on estimating the relevant moments of the distribution of purely axial displacements, resting assured that these bounds are unaffected by any small radial displacement. We will therefore set $\alpha=z$ and drop the directional subscript from all the forthcoming expressions, reprising the shorthand notation $\theta_l \equiv \theta_{l_z}$ to indicate axial moments.}

Since in this paper we focus on the problem of estimating the surface roughness encoded in the standard deviation (\ref{sigma})
$$
\sigma \equiv \tilde{\theta}_2 =\sqrt{\theta_2-\theta_1^2}, 
$$ 
we will ignore higher-order root central moments $\tilde\theta_{l>2}$ (such as those characterizing skewness and kurtosis). Using \eqref{eq:inverseQFI.theta} and 
\begin{equation}
\frac{\partial\tilde{\theta}_2}{\partial \theta_1 }= -\frac{\theta_1}{\tilde{\theta}_2}, \quad
\frac{\partial\tilde{\theta}_2}{\partial \theta_2 } = \frac{1}{2\tilde{\theta_2}},
\end{equation}
we compute the estimation error for the standard deviation as 
\begin{eqnarray}
\label{M2QFI}
    \mathcal{V}^{\rm Q}_{\sigma}=\frac{1}{4\Delta_0G^2} .   
\end{eqnarray}

Equations~(\ref{M1QFI}) and (\ref{M2QFI}) are the first main results of this paper. They show that, in the limit of small (sub-diffraction) displacements, the ultimate errors for the estimation of both centroid (mean height) and axial standard deviation (root-mean-square roughness) of a distribution of $N$ weak incoherent sources are {\it constant} and equal to $1/(4\Delta_0 G^2)$. Explicitly, given that for a Gaussian beam (\ref{Image plane PSF}) the variance of the generator $G$ is given by $\Delta_0 G^2=1/(4z_R^2)$, the errors for both the mean height and the  roughness parameter can be written more practically in terms of the Rayleigh range as 
\begin{equation}\label{gaussianresults}
\mathcal{V}^{\rm Q}_{\mu}=\mathcal{V}^{\rm Q}_{\sigma}=z_R^2.
\end{equation}

 Before proceeding onto the next section we note that these results nicely reproduce and generalize other results in the literature, computed in special cases using fewer sources. For example, in \cite{PhysRevLett.122.140505}, the QFI for the estimation of the axial separation of two incoherent sources $|z_2-z_1|$ is found to be $\Delta_0 G^2$ (see also \cite{YuPrasad2019,PrasadYu2019,Wang21,fiderer2021general,jernejgeorge,lvovsky2026passive}). Transforming this result to our coordinate system, where for two sources the axial standard deviation would be defined as $\sigma=|z_2-z_1|/2$, we recover $\mathcal{K}(\sigma)=4\Delta_0 G^2$. Here, we have shown that this expression holds for an arbitrary number $N$ of sources in the sub-diffraction limit. Our analysis also complements existing studies focused on estimating moments of the radial distribution (in the $x$ and/or $y$ directions) of non-axially displaced incoherent point sources \cite{Ang17,PhysRevA.99.013808,Tsang_2017,PhysRevResearch.1.033006}. Although our model considered a finite number of sources with known intensities, we found that in the limit of small size distributions, the estimation error for standard deviation does not depend on the number of sources and their positions. Since more general (continuous) distributions can be approximated with discrete ones by binning, this suggests that the same error rate should hold in general.  
 
 Now, equipped with our fundamental bound on the achievable information, let us see if we can reach it.

\section{Achieving the limits: \texorpdfstring{\\}{} Measurement Techniques}\label{SectionResults}

\red{In the remainder of this paper, we will specialize to the purely axial source distribution model illustrated in Figure~\ref{FigureB}. On the one hand, this is theoretically justified by the observation, obtained in the previous section, that the ultimate limits on estimating the axial moments in the sub-diffraction regime are not dependent on the distribution of the radial displacements. On the other hand, this case holds practical relevance in its own right for characterization techniques reliant on depth resolution, including laser interferometry for measuring microchannel depth or thin film thickness \cite{kolekar2012depth,park2024review}, as well as confocal photoluminescence for depth profiling or mapping depth distributions of defects in semiconductors \cite{maier2024modeling,song2023interpretation}.}

\subsection{Failure of Direct Imaging}\label{secfail}

 We will now show that direct imaging does not achieve the QFI for sub-diffraction axial moments estimation. For this, we use the methods of semi-parametric estimation \cite{PhysRevResearch.1.033006,PhysRevX.10.031023}, reviewed in section~\ref{SubSectionSemiP}. 
 Specifically, we will identify an efficient influence function $\varphi$ and use it to compute the bound \eqref{GHBB}.
This can be achieved by relating the object plane moments 
defined by Eq.~\eqref{object plane moment} to the moments in the image plane. According to \eqref{Reduced f}, the probability density associated to the direct imaging measurement is 
\begin{equation}
\label{GeneralDIIntensity}
    f_F(\boldsymbol{r})=\int|\psi(\boldsymbol{r},z)|^2 F(dz).
\end{equation}
where $F$ is the unknown axial source distribution. Note that while in section~\ref{SectionComputingQFI} we assumed $F$ to be supported on a finite number of points, 
in this section we do not make such restrictive assumptions, so $F$ will belong to a large (non-parametric) class of distributions. 

Since $f_F(\boldsymbol{r}) = f_F(r)$ is radially symmetric, it suffices to consider its even moments $\{\phi_{2j}\}_{j\geq 1}$ given by
\begin{equation}\label{phijmoments}
    \phi_{2j}=\int r^{2j} f_F(\boldsymbol{r})d\boldsymbol{r}.
\end{equation}
%

By expanding $H(\boldsymbol{r};z)=|\psi(\boldsymbol{r},z)|^2$ from Eq.~(\ref{Reduced H}) in a power series, we rewrite $\phi_{2j}$ as
\begin{eqnarray}
\label{DIexplicitRelatingThetaToPhi}
    \phi_{2j}&=&\sum_{k}\frac{1}{(2k)!}\int r^{2j}\left.\frac{\partial^{2k}|\psi(\boldsymbol{r},z)|^2}{\partial z^{2k}}\right\vert_{z=0}
    d\boldsymbol{r}\int z^{2k}dF(z) \nonumber \\
    &=&\sum_{k}C_{jk}\theta_{2k},
\end{eqnarray}
where we note that the coefficients of the odd moments $\theta_{2k+1}$ are zero and 
the matrix $C$ has elements given by
\begin{equation}
\label{DIexplicitC}
    [C]_{jk}=\frac{1}{(2k)!}\int r^{2j}\left.\frac{\partial^{2k}|\psi(\boldsymbol{r},z)|^2}{\partial z^{2k}}\right\vert_{z=0}d\boldsymbol{r}.
\end{equation}

The explicit form of $C$ is computed in Appendix \ref{AppednixBMatrixProof} as the lower triangular matrix with elements
\begin{equation}
\label{DICmat}
    [C]_{ij}=\frac{i!}{2^iz_R^{2j}}\binom{i}{j}.
\end{equation}
and its inverse is the upper triangular matrix with elements
\begin{equation}
\label{DICinvMat}
    [C^{-1}]_{ij}=(-1)^{i-j}\binom{i}{j}\frac{2^j}{j!}z_R^{2i}.
\end{equation}

Equation~(\ref{DIexplicitRelatingThetaToPhi}) can be expressed in vectorial form as $\boldsymbol{\phi}=C\boldsymbol{\theta}$ or, by taking inverses, 
    $\boldsymbol{\theta}=C^{-1}\boldsymbol{\phi}$. 
    Let $\beta =\beta(F)$ be a one-dimensional parameter of interest and let us assume that $\beta$ is of the form $\beta :=\boldsymbol{\partial\beta}^T\boldsymbol{\theta}$ for some fixed vector $\boldsymbol{\partial\beta}$. Let 
    $F_0$ be a reference source distribution, with $\beta_0 = \beta(F_0), \boldsymbol{\theta}_0 = \boldsymbol{\theta}(F_0)$, 
    $\boldsymbol{\phi}_0 = \boldsymbol{\phi}(F_0)$. Then
    $
    \beta =\beta_0 + \boldsymbol{\partial\beta}^T (\boldsymbol{\theta} -\boldsymbol{\theta}_0)
    $
and
\begin{equation}
    \beta=\beta_0 + (\boldsymbol{\partial\beta})^TC^{-1}( \boldsymbol{\phi} -\boldsymbol{\phi}_0) .
\end{equation}
Using (\ref{phijmoments}) we have then
\begin{eqnarray*}
\beta &=& \beta_0+
\int (f(\boldsymbol{r})  -f_0(\boldsymbol{r}) )d\boldsymbol{r} 
(\boldsymbol{\partial\beta})^T
C^{-1} \mathfrak{R}
\end{eqnarray*}
where $\mathfrak{R}$  is the vector with components $\mathfrak{R}_j:= r^{2j}$.
Therefore we conclude that
\begin{equation}\label{varphiDI}
\varphi= (\boldsymbol{\partial\beta})^TC^{-1} \mathfrak{R} - (\boldsymbol{\partial\beta})^TC^{-1}\boldsymbol{\phi}_0
\end{equation}
is an IF for the parameter $\beta$ and thus, by Eq.~(\ref{HB0}), the corresponding asymptotic scaled error in estimating $\beta$ is 
\begin{equation}
\label{DICRBGeneralForm}
\lVert\varphi\rVert^2_{F_0}=
\mathbb{E}_{F_0}[\varphi^2]=(\boldsymbol{\partial\beta})^TC^{-1} U C^{-T}\boldsymbol{\partial\beta},
\end{equation}
where $U$ is a matrix with elements
\begin{eqnarray}
\label{eq:U}
    [U]_{ij}&=&\int r^{2(i+j)}f_0(\boldsymbol{r})d\boldsymbol{r} - 
    \phi_{0,2i}\phi_{0,2j} \nonumber\\
    &=& \phi_{0,2(i+j)} -\phi_{0,2i}\phi_{0,2j}.
\end{eqnarray}

Similarly to \cite{PhysRevResearch.1.033006} it can be argued that $\varphi$ is an efficient IF, which means that the error bound in (\ref{DICRBGeneralForm}) cannot be improved. Intuitively, this is because $\varphi$ belongs to the subspace 
 of rotationally invariant functions in 
$L^{2}(\mathbb{R}^2, f_0)$, which coincides with the co-tangent space $\mathcal{T}$ of score functions. 
For more details and specific technical assumptions for which this result holds, we refer to Appendix \ref{AppendixDIIFefficiency}.

From \eqref{DICinvMat} and \eqref{eq:U} we obtain that
$\mathcal{V}^{\rm DI}_\theta = C^{-1}UC^{-T}$ is given by 
\begin{eqnarray}
\label{DICRBIntegralForm}
    [\mathcal{V}^{\rm DI}_{\boldsymbol{\theta}}]_{ij}
    &=&
    \sum_{n=0}^{i}\sum_{m=0}^{j}\binom{i}{n}\binom{j}{m}\frac{(-2)^{n+m}(-z_R^2)^{i+j}}{n!m!}
    \nonumber\\
    &\times& 
    (\phi_{0,2(n+m)}-\phi_{0,2n}\phi_{0,2m}).
\end{eqnarray}
and the optimal error for $\beta$ is 
\begin{equation}
    \mathcal{V}^{\rm DI}_\beta = 
    (\boldsymbol{\partial\beta})^T
    \mathcal{V}^{\rm DI}_{\boldsymbol{\theta}}
    \boldsymbol{\partial\beta}.
\end{equation}
In particular, the errors for estimating the even moments $\theta_{2j}$ are nonzero even in the limit 
of small sources spread due to the non-trivial value of the contributions $\phi_{0,2(n+m)}-\phi_{0,2n}\phi_{0,2m}$ computed with respect to the Gaussian density $f_0$ of the PSF. 


The above argument can be extended to non-linear functionals 
$\beta$ in which case $\boldsymbol{\partial}\boldsymbol{\beta}$ is the vector of partial derivatives at 
$\boldsymbol{\theta}_0 = \boldsymbol{\theta}(F_0)$. Since we focus on roughness, we set the first moment $\theta_1$ to zero and consider $\beta\equiv\sqrt{\theta_2}=\sigma$. 
Then
\begin{equation}
\label{dbeta}
    \boldsymbol{\partial\beta}_i=\frac{\delta_{1i}}{2\sqrt{\theta_2}},
\end{equation}
\red{with $i=0,1,2,\ldots$ spanning the even moments $\theta_{2i}$.}
The result of Eq.~(\ref{DICRBGeneralForm}) is that the CRB element for the estimation of the second root central moment, a.k.a.~roughness parameter, is (see Appendix \ref{AppendixDICRB} for the derivation)
\begin{eqnarray}
    \mathcal{V}^{\rm DI}_{\sigma}=
    \frac{z_R^4}{\sigma^2}\left(\phi_{0,4}-\phi^2_{0,2}\right).
\end{eqnarray}
In the sub-diffraction limit, our region of interest, it is assumed that $\Delta \ll 1$ and thus the standard deviation $\sigma=\sqrt{\theta_2} =\mathcal{O} (\Delta)\ll 1$ while 
$\phi_{0,j}$ converge to the moments of the Gaussian PSF of a single source $\psi(\boldsymbol{r},0)$. The result is a diverging CRB,
\begin{eqnarray}
    \lim_{\Delta\to 0}\mathcal{V}^{\rm DI}_{\sigma}
    =\infty.
\end{eqnarray}
Summarizing, in this section we have shown that direct imaging, incapable of estimating odd-ordered moments (including mean height), is also incapable of estimating the roughness (second root central moment) of an arbitrary distribution of axially displaced point sources below the Rayleigh limit.

\subsection{Triumph of SPADE}

Having shown that direct imaging fails to estimate surface roughness or, more precisely, the standard deviation of the axial distribution of incoherent point sources, 
we are motivated to search for a measurement scheme that is optimal for its estimation. 
In this section we show that a measurement strategy based on SPADE \cite{PhysRevX.6.031033,PhysRevA.99.012305} does have this property. We will begin this analysis by proceeding similarly to the previous section, but we will instead attempt to describe the vector $\boldsymbol{f} = [f_q]_{q\geq 0}$ of (normalized) intensities in the transverse mode basis 
linearly in terms of the object distribution moments as
\begin{equation}
\label{SPADEfInTermsOfTheta}
    \boldsymbol{f}=W \boldsymbol{\theta}.
\end{equation}

\begin{figure}[b]
\centering
\includegraphics[width=\columnwidth]{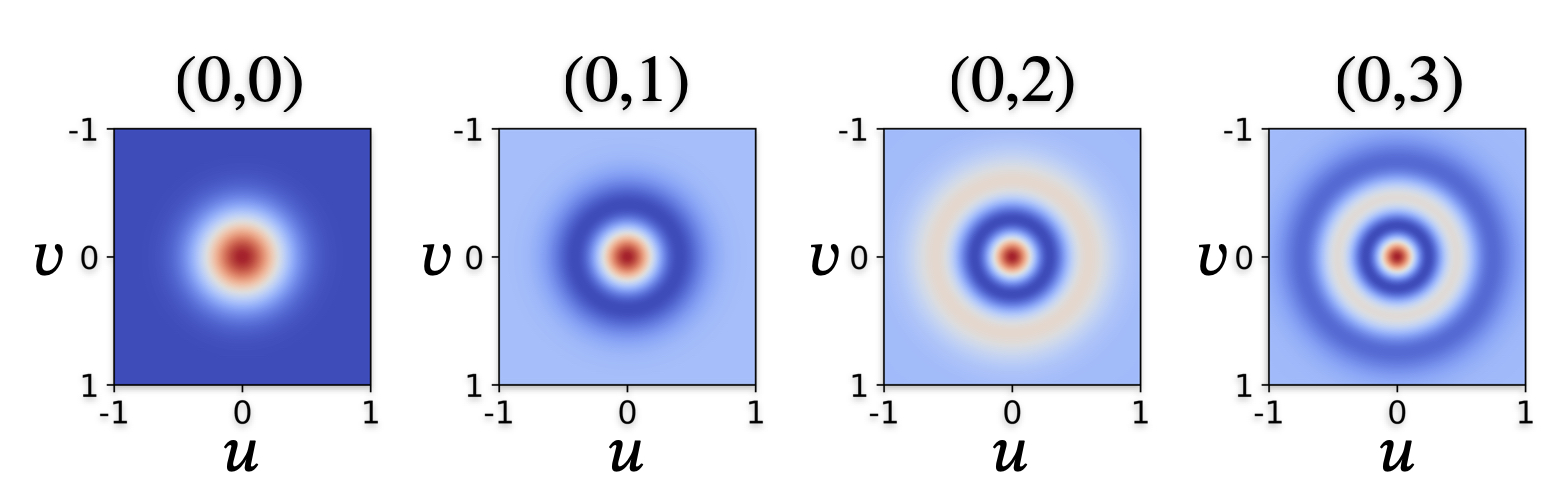}
\caption{Intensity plots of the first four radially symmetric Laguerre-Gauss modes for $\ell=0,p=0,1,2,3$ in the \red{$(u,\varv)$ image plane} centered at $(0,0)$. The $\ell=0,p=0$ mode is simply the PSF of an emitter in focus.}
\label{LGmodes}
\end{figure}

Due to our ``rabbit hole'' object being modelled as a distribution of sources displaced axially around the focus, we choose to use the Laguerre-Gauss (LG) basis (Figure~\ref{LGmodes}) as this has been shown to be optimal for two-source axial separation estimation \cite{Zhou:19}. 
The LG modes in the image plane are given in their normalized form by
\begin{eqnarray}\label{LGmodesImagePlane}
    \phi_q(\boldsymbol{r})=\sqrt{\frac{2}{\pi \omega_0^2}}L_q\left(\frac{2r^2}{\omega_0^2}\right)e^{\frac{-r^2}{\omega_0^2}}.
\end{eqnarray}
where $L_q(x)$ are the Laguerre polynomials, for $q\geq 0$. Due to our lack of radially displaced sources, our distribution is rotationally invariant about the optical axis and it suffices to consider the reduced LG basis which depends only on the radial index $q$. Equation~(\ref{Reduced f}) thus becomes
\begin{equation}
\label{SPADEf(q)}
    f_q=\int H(q;z)dF(z),
\end{equation}
where
\begin{equation}
\label{SPADEH}
    H(q;z)=|\langle\phi_q
    |\psi_{z}\rangle|^2,
\end{equation}
and $\ket{\phi_q
}$ represents the state of a photon in the mode $\phi_q(\boldsymbol{r})$,
\begin{equation}
    \ket{\phi_q
    }=\int\phi_q(\boldsymbol{r})\ket{\boldsymbol{r}}d\boldsymbol{r}.
\end{equation}
Given the Gaussian beam (\ref{Reduced Image plane PSF}), Eq.~(\ref{SPADEH}) is derived in Appendix \ref{AppednixCMatrixProof} to be
\begin{equation}
\label{SPADEHCompact}
    H(q;z)=\frac{4z_R^2z^{2q}}{(z^2+4z_R^2)^{q+1}}.
\end{equation}
By expanding in terms of $z$ we obtain
\begin{equation}
\label{SPADEHexplicit}
    H(q;z)=\sum^{\infty}_{k=0}\frac{(-1)^{k-q}}{(k-q)!}\frac{k!}{q!}\left(\frac{z}{2z_R}\right)^{2k}.
\end{equation}
This allows us to expand Eq.~(\ref{SPADEfInTermsOfTheta}) in components as 
\begin{equation}
\label{SPADEintensity}
f_q=\sum_{k=0}^{\infty}W_{qk}\theta_{2k},
\end{equation}
where $W$ is an upper triangular matrix with elements
\begin{equation}
\label{SPADECmatrixExplicit}
    [W]_{qk}=\frac{(-1)^{k-q}}{(2z_R)^{2k}}\binom{k}{q}.
\end{equation}
We note from the above expressions that, similarly to the case of direct imaging, only even-ordered moments $\theta_{2k}$ are recoverable by means of SPADE in the LG basis \footnote{In order to recover the odd-ordered moments, techniques such as measuring in the interferometric transverse electromagnetic basis would be necessary \cite{Tsang_2017}, but we do not do worry about that here.}. Inverting Eq.~(\ref{SPADEfInTermsOfTheta}) and expressing the parameter of interest 
as 
$\beta=(\boldsymbol{\partial\beta})^T\boldsymbol{\theta}$ where $\boldsymbol{\theta}$ is a vector consisting of only the even ordered moments, we find that
\begin{equation}
\label{SPADEbetaGeneral}
    \beta=\beta_0+ 
    (\boldsymbol{\partial\beta})^T(\boldsymbol{\theta} -\boldsymbol{\theta}_0)=\beta_0 +(\boldsymbol{\partial\beta})^T W^{-1}(\boldsymbol{f}-\boldsymbol{f}_0) ,
\end{equation}
where $\beta_0 = \beta(F_0), \boldsymbol{\theta}_0 = \boldsymbol{\theta}(F_0)$, 
    $\boldsymbol{f}_0 = \boldsymbol{f}(F_0)$, and the lower triangular matrix $W^{-1}$ is the inverse of the matrix in Eq.~(\ref{SPADECmatrixExplicit}), whose elements are (see Appendix \ref{AppednixCMatrixProof} for details) 
\begin{equation}
\label{SPADECinv}
    [W^{-1}]_{kq}=(2z_R)^{2k}\binom{q}{k}.
\end{equation}
 From Eq.~(\ref{SPADEbetaGeneral}) we identify an IF 
\begin{equation}
\label{SPADEIF}
    \varphi= (\boldsymbol{\partial\beta})^TW^{-1} \mathfrak{N}  - (\boldsymbol{\partial\beta})^T W^{-1}\boldsymbol{f}_0  .
\end{equation}
where $\mathfrak{N}$ is a vector with components $\mathfrak{N}_q= \delta_{q,Q}$ encoding the label of the measurement outcome $Q\in \mathbb{N}$.

The optimal error in estimating $\beta$ is given once again by the expectation value of the square of the efficient IF (see Appendix \ref{AppendixSPADEIFefficiency} for details and assumptions on $F_0$) , 
\begin{equation}
\label{SPADECRB}
\mathcal{V}^{\rm LG}_\beta =   \lVert\varphi\rVert_{F_0}^2=\mathbb{E}_{F_0}[\varphi^2]=(\boldsymbol{\partial\beta})^TW^{-1}D(W^{-1})^T\boldsymbol{\partial\beta} 
\end{equation}
where $D$ is a matrix with elements 
\begin{equation}
\label{SPADED}
[D]_{qp}=\delta_{qp}f_q - f_q f_p.
\end{equation} 
Evaluating the matrix $\mathcal{V}^{\text{LG}}_{\boldsymbol{\theta}}:= W^{-1}D(W^{-1})^T$ 
with our matrix $W^{-1}$ (\ref{SPADECinv}) and intensity $f_q$ (\ref{SPADEf(q)}), yields a matrix with elements
\begin{equation}
\label{SPADEgeneralCRB}
    [\mathcal{V}^{\text{LG}}_{\boldsymbol{\theta}}]_{ij}=\sum_{k=0}^{\min\{i,j\}}\binom{i+j-k}{j}\binom{j}{k}(2z_R)^{2k}\theta_{2(i+j-k)}-\theta_{2i}\theta_{2j}.
\end{equation}
Details of the derivation can be found in Appendix \ref{AppendixCRBSPADE}. Subsequently, for the estimation of $\sigma=\sqrt{\theta_2}$, application of Eqs.~(\ref{SPADECRB}) and (\ref{SPADEgeneralCRB}) with Eq.~(\ref{dbeta}) yields
\begin{eqnarray}
    \mathcal{V}^{\text{LG}}_{\sigma}=z_R^2+\frac{\theta_4}{2\theta_2}-\frac{\theta_2}{4}.
\end{eqnarray}
For a source distribution of size $\Delta$ we have 
$\theta_4\leq \Delta^2 \theta_2$ which decreases faster than $\theta_2=\mathcal{O}(\Delta^2)$, so in the limit of a very smooth surface patch (or a very shallow hole, more appropriately) $\Delta \ll 1$, we find
\begin{equation}
    \mathcal{V}^{\text{LG}}_{\sigma}=z_R^2
\end{equation}
This is equal to the ultimate precision limit $\mathcal{V}^Q_\sigma$ derived in \eqref{M2QFI}.
Our results thus show that SPADE performed in the LG basis is an optimal measurement technique for the estimation of axial roughness in the sub-diffraction limit.

\red{The metrological enhancement provided by SPADE stems from the fact that, as discussed in Section~\ref{SectionImagingTheory}, axial source displacements deform the optical field before they significantly alter the image intensity. In the sub-diffraction regime, neighbouring sources produce almost indistinguishable intensity profiles, rendering direct imaging insensitive to small variations in the source distribution and, consequently, to the roughness parameter. These axial deformations are instead naturally encoded in the LG mode structure of the field. By measuring the occupation probabilities of these modes, SPADE converts otherwise inaccessible information about the source distribution into observable intensity signals.} 

\subsection{\red{Measurement imperfections}}
\red{
Up to this point we have compared direct imaging and SPADE under idealized conditions. In practice, however, any implementation of SPADE is subject to experimental imperfections, such as mode crosstalk, optical losses, imperfect mode sorting, finite mode truncation, and misalignment between the imaging system and the mode sorter. Assessing the impact of these and other imperfections is important to determine the practical viability of quantum-inspired imaging and surface metrology schemes \cite{lvovsky2026passive}. Among the mentioned sources of error, misalignment represents one of the most common and experimentally relevant occurrences. In this section we therefore investigate the effect of axial misalignment between the incoming optical field and the SPADE apparatus. As a simple yet representative model, we numerically simulate the estimation of a discrete distribution of three axially displaced incoherent emitters, as illustrated in Figure~\ref{MisalignementModel}, and evaluate the resulting impact on roughness estimation.

\begin{figure}[ht]
\centering
\includegraphics[width=0.8\columnwidth]{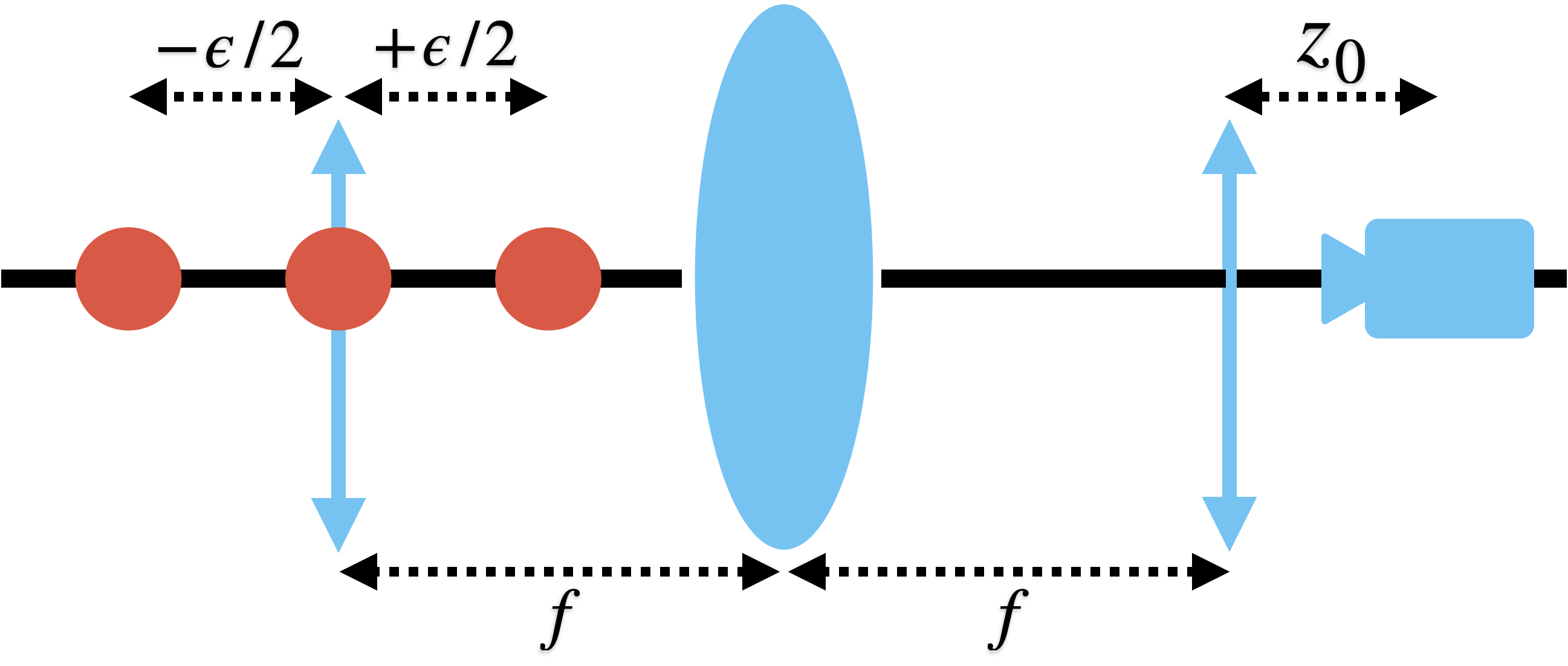}
\caption{\red{Schematic model of three sources with mode sorter misalignment. The three sources are placed at $-\epsilon/2$, $0$ and $+\epsilon/2$ relative to the focal point of the lens with focal length $f$. These sources form an axially distributed object with feature depth $\Delta=\epsilon$. The mode sorter is then displaced from the focal point on the other side of the lens by an amount $z_0$. Each of these displacements are in units of the Rayleigh range $z_R$.}}
\label{MisalignementModel}
\end{figure}

\red{We consider three sources distributed along the optical axis at coordinates $(-\epsilon/2,0,\epsilon/2)$, forming an object with feature depth $\Delta = \epsilon$.
We then consider a simple model of axial misalignment in which the optical axis of the mode sorter is displaced from the centroid $\theta_1=0$ of the object distribution by an amount $z_0$. Equivalently, the LG basis used for SPADE measurements is centered at an axial position offset from the nominal focal plane. Hence, the measurement modes are axially displaced by an amount $z_0$ and given by
\begin{eqnarray}
    \ket{\phi_{q,z_0}}=e^{-iz_0G}\ket{\phi_q}.
\end{eqnarray}
Since the same propagation generator $G$ governs both source evolution and mode displacement, this offset can be absorbed into an effective shift of the source coordinates, $z \rightarrow z-z_0$. Consequently, the probability (\ref{SPADEH}) of detecting a photon in the mode $\phi_q$, when using a misaligned mode sorter, is given by
\begin{eqnarray}
    H(q;z-z_0)=|\bra{\phi_q}e^{-i(z-z_0)G}\ket{\psi_{0}}|^2=|\langle \phi_q|\psi_{z-z_0}\rangle|^2.\ \ \ \ 
\end{eqnarray}}

\red{In Figure~\ref{SPADEcfiMisalignement} we plot the classical Fisher information for estimating the root-mean-square roughness  $\sigma$ using SPADE in the LG basis for different values of the axial misalignment $z_0$, and compare it with direct imaging. The curves are shown as a function of the normalized feature depth $\Delta/z_R$. Note that the probability of measuring $\geq4$ modes is of the order of $10^{-4}$, therefore we have limited our simulations to SPADE measurements using up to $3$ LG modes only.}

\begin{figure}[ht]
\centering
\includegraphics[width=\columnwidth]{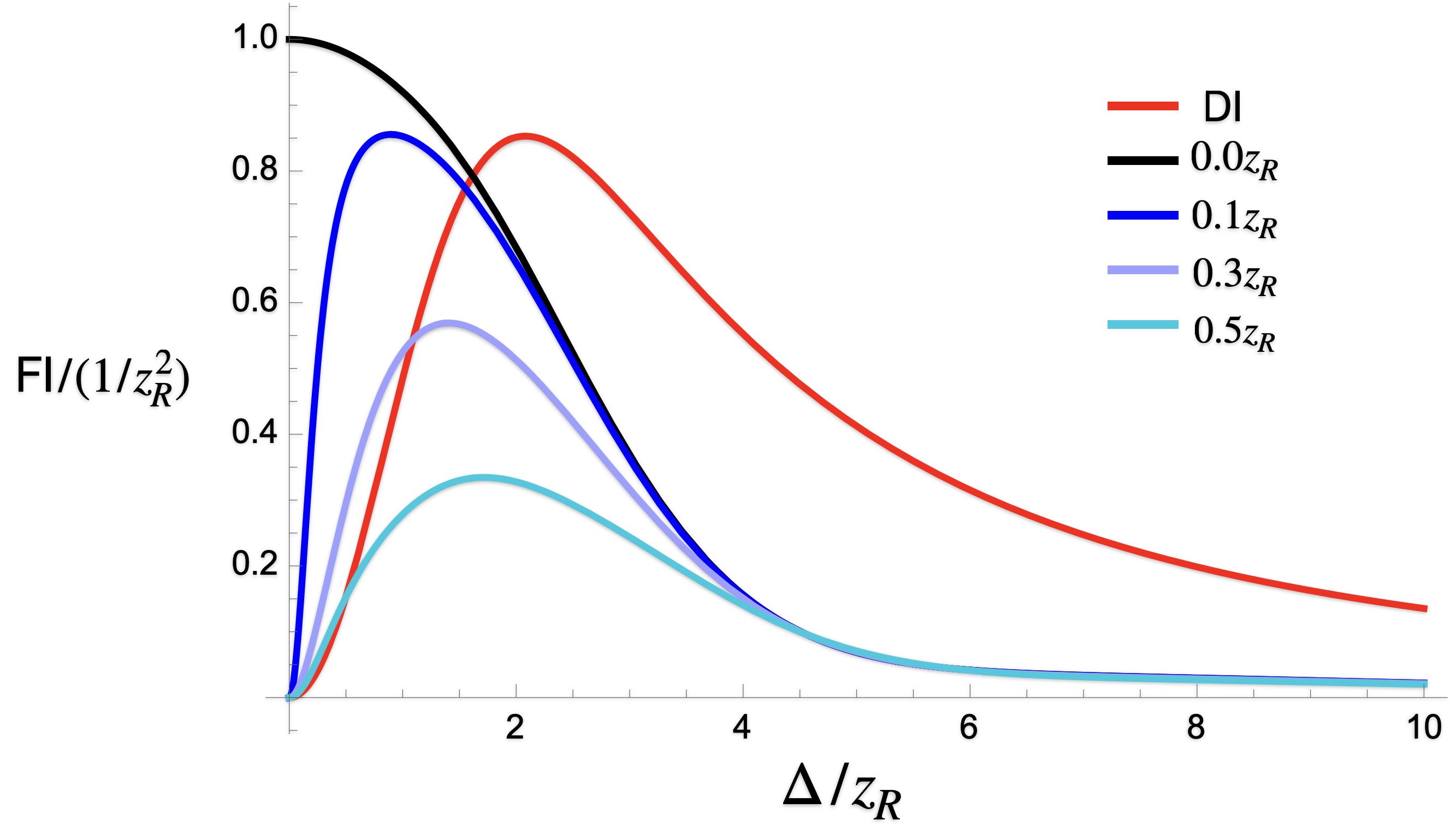}
\caption{\red{Classical Fisher information (normalized to the quantum Fisher information bound) for the estimation of root-mean-square roughness as a function of the feature depth (normalized to the Rayleigh range), plotted for the source distribution of Figure~\ref{MisalignementModel}, when using direct imaging (DI) versus SPADE detectors for different values of the axial misalignment $z_0/z_R=0, 0.1, 0.3, 0.5$.}}
\label{SPADEcfiMisalignement}
\end{figure}
}

\red{Figure~\ref{SPADEcfiMisalignement} shows that the performance of SPADE gradually degrades under axial misalignment. While any non-zero offset $z_0$ prevents the classical Fisher information of SPADE measurements from saturating the quantum Fisher information bound in the limit $\Delta \rightarrow 0$, SPADE continues to outperform direct imaging throughout a substantial portion of the sub-diffraction regime.  This behaviour mirrors that observed in radial mode-sorting schemes and in experimental demonstrations of axial two-source separation \cite{Zhou:19}, where precise registration of the measurement basis with the object is essential for optimal performance. In the radial case, this difficulty can be overcome by first estimating the centroid via direct imaging, which remains optimal for that task. By contrast, as we have shown in section~\ref{secfail}, for purely axial distributions direct imaging itself fails to provide reliable information about the first moment in the sub-diffraction regime, making the calibration of axial SPADE measurements intrinsically more challenging and motivating the development of robust alignment and adaptive measurement strategies. A systematic investigation of additional imperfections affecting mode sorting technology, together with practical mitigation and calibration strategies, represents an important direction for future work.}

\section{Discussion}\label{SectionDiscussion}

\red{Within this paper we have developed a rigorous metrological framework for the quantum-inspired estimation of surface roughness in near-smooth samples modelled as distributions of incoherent optical point sources. By treating the source positions as parameters encoded in the optical field, we have derived the ultimate quantum precision bounds for the estimation of spatial moments of arbitrary discrete three-dimensional source distributions in the sub-diffraction regime. This establishes a quantitative benchmark for how much information about the source can, in principle, be extracted from the optical field before any particular measurement scheme is chosen.}

\red{As a particularly relevant case, we have then specialized the analysis to purely axial source distributions, corresponding to depth variations of a small surface patch. This regime captures the essential metrological challenge in applications where axial resolution is the limiting factor, including optical surface inspection, thin-film and microstructure characterization, semiconductor defect profiling, and precision manufacturing metrology \cite{kolekar2012depth,park2024review,maier2024modeling,song2023interpretation,catalucci2022optical}.}

We have shown, through the elegant and powerful language of quantum metrology, that the fundamental error in estimating the roughness (standard deviation) of the purely axial distribution of an arbitrary number of point sources is finite and achievable. Namely, this can be achieved by demultiplexing the beam into a basis spanned by Laguerre-Gauss modes, followed by intensity measurements. Moreover we have shown that direct imaging fails to reach this bound and that, as the roughness becomes smaller, the information attainable about it with such a measurement technique goes to zero.  \red{We have also considered the effect of measurement imperfections such as axial misalignment between the incoming field and the demultiplexing apparatus. We have found that the performance of Laguerre-Gauss demultiplexing is degraded by misalignment, but remains superior to direct imaging in the deep sub-diffraction regime.} 

\red{Several interesting extensions remain open. First of all, building on existing investigations of radial source distributions  \cite{Ang17,PhysRevA.99.013808,Tsang_2017,PhysRevResearch.1.033006,tsang2019resolving}, together with the axial distribution studied here, in future work we aim to investigate the joint estimation of multiple parameters of relevance characterizing a three-dimensional object. While we have shown that radial and axial moments estimation can be analyzed separately at the level of ultimate quantum bounds, it remains to determine which experimentally feasible measurements can approach optimal performance for several three-dimensional parameters simultaneously. In particular, detection strategies that are optimal for purely axial roughness might not be jointly optimal once transverse structure is included.}

Whilst our derivation of the {\it quantum} Fisher information and related bounds in this paper has been for a discrete distribution of sources, we would like to generalize this to a continuum to capture a more realistic description of surface profiles; thankfully, the semi-parametric estimation techniques --- here employed for the evaluation of {\it classical} estimation bounds for the different measurement strategies --- will continue to apply in such a setting, given that the parameters of interest are limited to one or two moments, despite the infinite dimensionality of the nuisance coordinate space.  

Finally, some of the most powerful current surface analysis implementations utilize laser interferometry. As such, it can be desirable to look into a model consisting of {\em coherent} superpositions of sources and take into account the effects of illumination more in detail, as opposed to the incoherent light model adopted in this work.

\vspace{2mm}

\begin{acknowledgments}
We thank Mankei Tsang for useful discussions.
We acknowledge financial support from the Engineering and Physical Sciences Research Council (Grants No.~EP/T022140/1, EP/W028131/1, and EP/X010929/1). 
\end{acknowledgments}

\appendix

\section{Derivation of the {\it \bfseries C} Matrix for Direct Imaging}
\label{AppednixBMatrixProof}

In this section we prove that the matrix that linearly relates object moments and image moments given by Eq.~(\ref{DIexplicitC}) can be expressed in the more compact factorial form given by Eq.~(\ref{DICmat}). We can express (\ref{DIexplicitC}) as
\begin{equation}
\label{AppendixBDICmatInitial}
    [C]_{jk}=\frac{1}{(2k)!}\left. \frac{\partial^{2k}}{\partial z^{2k}}\int|\psi(\boldsymbol{r};z)|^2 r^{2j}d\boldsymbol{r} \right\vert_{z=0}.
\end{equation}
The integral in the expression above can be expressed explicitely in terms of the Gaussian beam (\ref{Reduced Image plane PSF}) as
\begin{equation}
I_j(z)=\frac{4}{\omega(z)^2}\int_0^{\infty} r^{2j+1}e^{\frac{-2r^2}{\omega(z)^2}}dr.
\end{equation}
Reparameterizing $r^2\rightarrow 2x$ we can rewrite this as
\begin{equation}
    I_j(z)=\frac{2}{\omega(z)^2}\int_0^{\infty}x^je^{\frac{-2x}{\omega(z)^2}}dx.
\end{equation}
This integral is well known \cite{IntegralTable} and can be evaluated as
\begin{equation}
    I_j(z)=\frac{j!\omega(z)^{2j}}{2^j}.
\end{equation}
The elements of the matrix $C$ are thus written as
\begin{equation}
    [C]_{jk}=\frac{j!}{2^j(2k)!}\left. \frac{\partial^{2k}}{\partial z^{2k}} \left(1+\frac{z^2}{z_R^2}\right)^{j} \right\vert_{z=0}.    
\end{equation}
This can be rewritten using the binomial theorem as
\begin{equation}
[C]_{jk}=\frac{j!}{2^j(2k)!}\sum_{l=0}z_R^{-2l}\binom{j}{l}\frac{(2l)!}{(2l-2k)!}\left. z^{2(l-k)} \right\vert_{z=0}. 
\end{equation}
Evaluated at $z=0$, the only surviving term is the one corresponding to $k=l$ and so the elements of the $C$ matrix reduce to the compact form,
\begin{equation}
    [C]_{jk}=\frac{j!}{z_R^{2k}2^j}\binom{j}{k}.
\end{equation}

Now we prove that the matrix given by Eq.~(\ref{DICinvMat}) is indeed the inverse of matrix (\ref{DICmat}). In order to show this, we must demonstrate that  $[CC^{-1}]_{ij}\equiv S_{ij}$ equals $\delta_{ij}$, where
\begin{equation}
    S_{ij}=(-1)^j2^{j-i}\frac{i!}{j!}\binom{i}{j}\sum_{m=0}^{\infty}(-1)^j\binom{i}{m}\binom{m}{j}.
\end{equation}
Using the known binomial identity,
\begin{equation}
\label{BinomialID}
    \binom{i}{m}\binom{m}{j}=\binom{i}{j}\binom{i-j}{m-j},
\end{equation}
one can rewrite $S_{ij}$ as
\begin{equation}
\label{S_ijReduced}
    S_{ij}=(-1)^j2^{j-i}\frac{i!}{j!}\binom{i}{j}^2\sum_{m=j}^{\infty}(-1)^j\binom{i-j}{m-j}.
\end{equation}
In the case of $i=j$, the expression simplifies to
\begin{equation}
    S_{jj}=(-1)^j\sum_{m=j}^{j}(-1)^j\binom{0}{m-j}=(-1)^j(-1)^j\binom{0}{0}=1.
\end{equation}
On the other hand for $i\neq j$, we can reparametrize $i-j\rightarrow k:k>0$ and $m-j \rightarrow l$ in order to rewrite (\ref{S_ijReduced}) as
\begin{equation}
    S_{kj}=\frac{(k+j)!}{j!}\binom{k+j}{j}2^{-k}\sum_{l=0}^{k>0}(-1)^l\binom{k}{l}.
\end{equation}
This final summation is proportional to the series expansion of $(1-x)^{k}$ with $x=1$ and is thus equal to zero $\forall k>0$. As such we can conclude that $S_{ij}=\delta_{ij}$ and (\ref{DICinvMat}) is indeed the inverse of (\ref{DICmat}).

\section{Direct Imaging Influence Function Efficiency}\label{AppendixDIIFefficiency}
In this section we will prove that the IF $\varphi\in\mathcal{H}_{F_0}$ given by (\ref{varphiDI}) is efficient.
We begin by defining the space of rotation invariant square-integrable functions on $\mathbb{R}^2$, that is, square-integrable with respect to the measure $f_0(r)=\int H(\boldsymbol{r};z)F_0(dz)$ representing the intensity at the reference parameter $\boldsymbol{\theta}_0$,
\begin{equation}
\begin{aligned}
    L^2_{\text{rad}}(f_0(r))=\bigg\{& g(\boldsymbol{r}):\mathbb{R}^2\rightarrow\mathbb{C}~ \big\lvert ~g(\boldsymbol{r})=g(\abs{\boldsymbol{r}}),\\
    &\int\abs{g(\boldsymbol{r})}^2 f_0(\boldsymbol{r})d\boldsymbol{r}<\infty\bigg\}.
\end{aligned}
\end{equation}
We then define the subspace of zero-mean functions as
\begin{equation}
L^2_{\text{rad},0}(f_0)=\bigg\{g(\boldsymbol{r})\in L^2_{\text{rad}}(f_0(r))\ \big\lvert \int g(\boldsymbol{r})f_0(\boldsymbol{r})d\boldsymbol{r}=0\bigg\}.
\end{equation}
Let us clarify here that $F_0$ is a reference measure on the measureable space $(\mathbb{R},\mathcal{A})$ where the sample space represents the positions of light sources. Throughout this work we are only concerned with estimating distributions with very small feature sizes. 
Thus we make the assumption that the support of $F_0$ is a compact subset of the set $A=[-a,a]\subset\mathbb{R}$ where $a\in(0,z_R)$. 
For technical reasons we also assume that the support of $F_0$ in $A$ contains an infinite number of points, where we recall that a point $x$ is in the support of $F_0$ if 
$F_0( (x-\epsilon, x+\epsilon) \cap A) >0$ for all $\epsilon>0$.

We define the statistical model $\mathcal{P}$ as the collection of distributions that are dominated by the reference distribution, $\mathcal{P}=\{F:F \ll F_0\}$. Given that our intensities are related to the source distribution $F(z)$ and this distribution is generally of an unknown form, we must assume an infinite-dimensional parameter $\boldsymbol{\theta}$ and thus we will instead consider a one-dimensional parametrization mapping a neighborhood of $t=0$ in $[0,\infty)$ to the statistical model $\mathcal{P}$, given by
\begin{eqnarray}\label{onepiece}
    t\mapsto F_t(z).
\end{eqnarray}
Explicitly we will use a normalized exponential family:
\begin{eqnarray}\label{normafamily}
    F_{t}(dz)=\frac{e^{t u(z)}F_0(dz)}{\int e^{t u(z)}F_0(dz)}.
\end{eqnarray}
Here $u(z)\in L^{\infty}(F_0)$ is any zero-mean (with respect to the measure $F_0(dz)$) measurable function. We define the expectation with respect to $F_0$ as $\mathbb{E}_{F_0}[u]=\int u(z) F_0(dz)$. 
Given the image intensity $f_t(r)=\int H(\boldsymbol{r};z)F_t(dz)$, we can write the scores as 
\begin{eqnarray}
    S_u(\boldsymbol{r})=\left.\frac{\partial}{\partial t}\right\vert_{t=0}\log f_t(\boldsymbol{r})=\frac{\int H(\boldsymbol{r};z)u(z)F_0(dz)}{f_0(\boldsymbol{r})}
\end{eqnarray}
Let us define the space of score functions as $\mathfrak{S} \equiv \{S_u(\boldsymbol{r})\}_u$. We aim to show that the span of this space, i.e., the co-tangent space $\mathcal{T}\equiv\text{span}\{\mathfrak{S}\}$, is dense in the subspace $L^2_{\text{rad},0}(f_0)$ of zero-mean square-integrable rotation-invariant functions. That is, we aim to show:
\begin{eqnarray}
    \overline{\mathcal{T}}=L^2_{\text{rad},0}(f_0).
\end{eqnarray}
We will do this by showing that the orthogonal complement $\mathcal{T}^\perp$ of the co-tangent space is a trivial subspace, i.e. $\mathcal{T}^\perp=\{0\}$. 
The key takeaway for us will be that an IF belonging to $L^2_{\text{rad},0}(f_0)$ will thus automatically belong to $\overline{\mathcal{T}}$ and therefore be efficient.

Let us begin by considering a function $g(\boldsymbol{r})\in L^2_{\text{rad},0}(f_0)$ that is orthogonal to the scores of each sub-model:
\begin{eqnarray}
    \langle g,S_u\rangle_{f_0}&= 
    \int g(\boldsymbol{r}) S_u(\boldsymbol{r}) f_0(\boldsymbol{r}) d\boldsymbol{r}=0.
\end{eqnarray}  
We first show that the function
$$
G(z):=\int g(\boldsymbol{r})H(\boldsymbol{r};z)d\boldsymbol{r}$$
is a well defined bounded function for $z\in A$. Indeed, for any  $z\in A$,
\begin{equation}
\label{eq:bounds.H}
c e^{-\frac{2r^2}{\omega_0^2}}\leq H(\boldsymbol{r};z) \leq 
C e^{-(1+\epsilon)\frac{r^2}{\omega_0^2}},
\end{equation}
    for some constants $c,C>0$ and a constant 
$\epsilon>0$ which depends on $a$. Therefore by using Cauchy-Schwarz
\begin{eqnarray}
\int |g(\boldsymbol{r})|H(\boldsymbol{r};z)d\boldsymbol{r}
&\leq& 
\int |g(\boldsymbol{r})| e^{-\frac{r^2}{\omega_0^2}} 
e^{\frac{-\epsilon r^2}{\omega_0^2}}d\boldsymbol{r}
\nonumber \\
&\leq &\tilde{C} \int |g(\boldsymbol{r})|^2 
e^{-\frac{2r^2}{\omega_0^2}}d\boldsymbol{r}
\\ \nonumber
&\leq &\tilde{C}\|g\|^2_{f_0}<\infty.
\end{eqnarray}
where we used that 
$f_0(\boldsymbol{r})\geq c e^{-\frac{2r^2}{\omega_0^2}}$, which follows from \eqref{eq:bounds.H}. Note that $G(z)$ has zero mean with respect to $F_0$.
Now we have
\begin{eqnarray}
\langle g, S_u\rangle_{_{f_0}} &=&
\iint 
\overline{g(\boldsymbol{r})} H(\boldsymbol{r},z) u(z) F_0(dz) d\boldsymbol{r}  \\
   \nonumber
   &=&
\int u(z) \overline{G(z)}F_0(dz) =
\langle G(z), u(z)\rangle_{F_0}=0.
\end{eqnarray}
Since $u(z)$ is arbitrary and $G$ has zero mean, this implies that
\begin{eqnarray}
\label{Gzeroae}
    G(z)=0,\,\,\,\,\,F_0-a.e. \,\,z\in\mathbb{R}.
\end{eqnarray}
 In order to extend the condition $G(z)=0$ to the entire interval $A$, we will show that $G(z)$ is analytic on $A$. We begin by noting that for the Gaussian direct imaging model,
\begin{eqnarray}
\label{HdiModel}
    H(\boldsymbol{r};z)=\frac{2}{\pi \omega(z)^2}e^{-\frac{2r^2}{\omega(z)^2}}.
\end{eqnarray}
Given that $g(\boldsymbol{r})$ is rotation invariant and subsequently defining $\abs{\boldsymbol{r}}^2\equiv t$ and $\zeta(t)\equiv g(\sqrt{t})$, we can write $G$ in terms of the Laplace transform,
\begin{equation}
\label{LaplaceTransform}
    G(z)=\lambda(z)\int^{\infty}_0\zeta(t)e^{-\lambda(z)t}dt=\lambda(z)\mathcal{L}\{\zeta\}(\lambda),
\end{equation}
where
\begin{equation}
\label{DIlambda}
    \lambda(z)=\frac{2}{\omega^2(z)} =\frac{2}{\omega_0^2\left(1+ \frac{z^2}{z_R^2}\right) }.
\end{equation}
Since $G(z)$ was shown to be finite, the Laplace transform  $\mathcal{L}\{\zeta\}(\lambda)$ exists for an interval of values for $\lambda$, which in particular means that it is analytic in a half plane. Since the support of $F_0$ is infinite and $G(z)$ is continuous and zero, $F_0-a.e.$, it follows that $G(z)$ is identically zero, and so is $g$.

Therefore, seeing as the only orthogonal element in $L^2_{\text{rad},0}(f_0)$ is the zero function, we conclude that the orthogonal complement of $\mathcal{T}$ is trivial and thus $\mathcal{T}$ is dense in $L^2_{\text{rad},0}(f_0)$. In conclusion any element in $L^2_{\text{rad},0}(f_0)$ can be arbitrarily well approximated by elements in the space of scores $\mathfrak{S}$ and thus an IF $\varphi\in L^2_{\text{rad},0}(f_0)$ implies that $\varphi$ is automatically efficient.

\section{Derivation of the Direct imaging CRB}\label{AppendixDICRB}
In this section we provide the derivation of the CRB for the estimation of the surface roughness by direct imaging. Given the CRB matrix for moments to be (\ref{DICRBIntegralForm}),
\begin{equation}
\begin{aligned}
    [\mathcal{V}^{\text{DI}}_{\boldsymbol{\theta}}]_{ij}&=
    \sum_{n=0}^{\infty}\sum_{m=0}^{\infty}\binom{i}{n}\binom{j}{m}\frac{(-2)^{n+m}(-z_R^2)^{i+j}}{n!m!}\\
        &\times\left(\int \boldsymbol{r}^{2(n+m)}f_0(\boldsymbol{r})d\boldsymbol{r}-\phi_{0,2n}\phi_{0,2m}\right),
    \end{aligned}
\end{equation}
and the vector of partial derivatives having elements
\begin{equation}
    \boldsymbol{\partial\beta}_i=\frac{\delta_{1i}}{2
    \sqrt{\theta_2}},
\end{equation}
we can write the error for the estimation of the surface roughness parameter $\beta \equiv \sigma$ as
\begin{eqnarray}
    \mathcal{V}^{\text{DI}}_\sigma &=&(\boldsymbol{\partial\beta})^TC^{-1}UC^{-T}\boldsymbol{\partial\beta} \nonumber\\
    &=&\frac{z_R^4}{4\theta_2}\sum_{n=0}\sum_{m=0}\binom{1}{n}\binom{1}{m}\frac{(-2)^{n+m}}{n!m!}\\
    &\times&\left(\int \boldsymbol{r}^{2(n+m)}f_0(\boldsymbol{r})d\boldsymbol{r}-\phi_{0,2n}\phi_{0,2m}\right). \nonumber
\end{eqnarray}
Writing out these sums explicitly, we find
\begin{equation}
\begin{aligned}
\mathcal{V}^{\text{DI}}_\sigma&=\frac{z_R^4}{4\theta_2}\left(\phi_{0,0}-\phi_{0,0}\phi_{0,0}\right)
    -\frac{z_R^4}{2\theta_2}\left(\phi_{0,2}-\phi_{0,2}\phi_{0,0}\right) \\
    &-\frac{z_R^4}{2\theta_2}\left(\phi_{0,2}-\phi_{0,0}\phi_{0,2}\right)
    +\frac{z_R^4}{\theta_2}\left(\phi_{0,4}-\phi_{0,2}\phi_{0,2}\right).
\end{aligned}
\end{equation}
The first three terms disappear and we are left with
\begin{eqnarray}
    \mathcal{V}^{\text{DI}}_\sigma=\frac{z_R^4}{\theta_2}\left(\phi_{0,4}-\phi_{0,2}\phi_{0,2}\right).
\end{eqnarray}
Due to the positivity of the variance, we can assume the difference appearing in this quantity is always positive.

\section{Derivation of the {\it \bfseries W} Matrix for SPADE}\label{AppednixCMatrixProof}
Using the Gaussian beam (\ref{Image plane PSF}) and the ${q}^{\rm{th}}$ LG mode (\ref{LGmodesImagePlane}) in the image plane, Eq.~(\ref{SPADEH}) can be written explicitely as
\begin{eqnarray}
    H(q;z)=\frac{16}{\omega(z)^2}\abs*{\int^{\infty}_0 re^{-ar^2}L_q(br^2)dr}^2,
\end{eqnarray}
where $b=2$ and
\begin{eqnarray}
    a=1+\frac{1}{\omega(z)^2}\left(1+\frac{iz}{z_R}\right).
\end{eqnarray}
This integral is known \cite{IntegralTable} and thus one finds
\begin{eqnarray}
    H(q;z)=\frac{4z_R^2z^{2q}}{(4z_R^2+z^2)^{q+1}}.
\end{eqnarray}
We now prove that the matrix given by Eq.~(\ref{SPADECinv}) with elements
\begin{equation}
    [W^{-1}]_{kq}=(2z_R)^{2k}\binom{q}{k},
\end{equation}
is the inverse of the matrix given by Eq.~(\ref{SPADECmatrixExplicit}) with elements
\begin{equation}
    [W]_{qk}=\frac{(-1)^{k-q}}{(2z_R)^{2k}}\binom{k}{q}.
\end{equation}
Similarly to what is done in Appendix \ref{AppednixBMatrixProof}, we must show that the elements of the product $WW^{-1}$ reduce to a Kronecker delta. As such for the elements $[WW^{-1}]_{ij}\equiv S_{ij}$ we must prove
\begin{equation}
    S_{ij}=\frac{i!}{j!}(-1)^i\sum_{m=0}^{\infty}\frac{(-1)^m}{(m-i)!(j-m)!}=\delta_{ij},
\end{equation}
We can rewrite the above expression as
\begin{equation}
\begin{aligned}
    S_{ij}&=\frac{i!}{j!}(-1)^i\sum_{m=0}^{\infty}\frac{(-1)^mm!m!i!j!}{(m-i)!(j-m)!m!m!i!j!} \\
    &=(-1)^i\sum_{m=0}^{\infty}(-1)^m\binom{j}{m}\binom{m}{i}.
\end{aligned}
\end{equation}
Again using (\ref{BinomialID}) we can write
\begin{equation}
\label{SSPADE}
    S_{ij}=(-1)^i\binom{j}{i}\sum_{m=0}^{\infty}(-1)^m\binom{j-i}{m-i}.
\end{equation}
For the case $j=i$ we can simplify this to
\begin{equation}
    S_{jj}=(-1)^j\sum_{m=j}^{j}(-1)^m\binom{0}{m-j}=(-1)^j(-1)^j\binom{0}{0}=1.
\end{equation}
Alternatively for the case $i\neq j$ we perform the transformation $j-i\rightarrow k:k>0$ and $m-i \rightarrow l$ such that (\ref{SSPADE}) becomes
\begin{equation}
    S_{ik}=\binom{i+k}{i}\sum_{l=0}^{k>0}(-1)^l\binom{k}{l}.
\end{equation}
This final summation is proportional to the series expansion of $(1-x)^{k}$ with $x=1$ and is thus equal to zero $\forall k>0$. As such we can conclude that $S_{ij}=\delta_{ij}$ and (\ref{SPADECinv}) is indeed the inverse of (\ref{SPADECmatrixExplicit}). 

\section{Derivation of the CRB for SPADE}\label{AppendixCRBSPADE}
In this section we derive the CRB for the estimation of moments (\ref{SPADEgeneralCRB}) using SPADE in the LG basis. Writing $\mathcal{V}^{\text{LG}}_{\boldsymbol{\theta}}:= W^{-1}D(W^{-1})^T$ with Eqs.~(\ref{SPADECinv}), (\ref{SPADED}) and (\ref{SPADEHCompact}) yields a matrix with elements
\begin{align}
    \label{VwithSeries}
    [\mathcal{V}^{\text{LG}}_{\boldsymbol{\theta}}]_{ij} &=(2z_R)^{2(i+j)}\bigg[\int\frac{(2z_R)^{2}}{(z^2+4z_R^2)}S^{(1)}_{ij}dF(z)\\
     &-\int\frac{(2z_R)^2}{z^2+4z_R^2}S_i^{(2)}dF(z)\int\frac{(2z_R)^2}{z^2+4z_R^2}S_j^{(2)}dF(z)\bigg], \nonumber
\end{align}
where
\begin{eqnarray}
    S_{ij}^{(1)}=\sum_{n=0}^{\infty}\binom{n}{i}\binom{n}{j}x^n, \quad
    S_i^{(2)}=\delta_{ij}S_j^{(2)}=\sum_{n=0}^{\infty}\binom{n}{i}x^n, \quad
\end{eqnarray}
and $x=z^2/(z^2+4z_R^2)$. Using the known binomial expansion,
\begin{eqnarray}
\label{BinomialExpansionFormula}
    \sum_{n=0}^{\infty}\binom{n}{i}x^n=\frac{x^i}{(1-x)^{i+1}},
\end{eqnarray}
we find 
\begin{eqnarray}
\label{S2series}
    S_i^{(2)}=\delta_{ij}S_j^{(2)}=\frac{z^{2i}(z^2+4z_R^2)}{(2z_R)^{2(i+1)}}.
\end{eqnarray}
To compute $S_{ij}^{(1)}$, we consider the useful binomial identity,
\begin{eqnarray}
    \binom{n}{i}\binom{n}{j}=\sum_{k=0}^{\min\{i,j\}}\frac{(i+j-k)!}{k!(i-k)!(j-k)!}\binom{n}{i+j-k}.
\end{eqnarray}
Multiplying through by $x^n$ and summing to infinity over $n$, we find
\begin{align}
S_{ij}^{(1)} &= \sum_{n=0}^{\infty}\binom{n}{i}\binom{n}{j}x^n \\
&= \sum_{k=0}^{\min\{i,j\}}\frac{(i+j-k)!}{k!(i-k)!(j-k)!}\nonumber
   \sum_{n=0}^{\infty}\binom{n}{i+j-k}x^n.
\end{align}
Reusing the identity (\ref{BinomialExpansionFormula}), we conclude that
\begin{equation}
\label{S1series}
    S_{ij}^{(1)}=\frac{x^{i+j}}{(1-x)^{i+j+1}}\sum_{k=0}^{\min\{i,j\}}\binom{i+j-k}{j}\binom{j}{k}\left(\frac{1-x}{x}\right)^{k}.
\end{equation}
Substituting (\ref{S1series}) and (\ref{S2series}) back into (\ref{VwithSeries}), we find
\begin{align}
    [\mathcal{V}^{\text{LG}}_{\boldsymbol{\theta}}]_{ij}=\sum_{k=0}^{\min\{i,j\}}\binom{i+j-k}{j}\binom{j}{k}(2z_R)^{2k}\theta_{2(i+j-k)}
    -\theta_{2i}\theta_{2j}.    
\end{align}
To reparameterize to the roughness parameter we compute
\begin{eqnarray}
    \mathcal{V}^{\rm LG}_\sigma =(\boldsymbol{\partial\beta})^T\mathcal{V}^{\text{LG}}_{\boldsymbol{\theta}}\boldsymbol{\partial\beta},
\end{eqnarray}
where the vector of partial derivatives is given by
\begin{eqnarray}
    (\boldsymbol{\partial\beta})^T=\left(0,\frac{1}{2\sqrt{\theta_2}},0,\dots\right)^T.
\end{eqnarray}

\section{SPADE Influence Function Efficiency}\label{AppendixSPADEIFefficiency}
In this section we will prove that the IF $\varphi\in\mathcal{H}_{F_0}$ given by (\ref{SPADEIF}) is efficient. We will proceed in a similar manner to Appendix (\ref{AppendixDIIFefficiency}) in that we will show the span of the space of scores for the SPADE model to be dense in the Hilbert space containing the SPADE model intensities. In the SPADE model, single photons are measured in the Laguerre-Gauss basis (\ref{LGmodesImagePlane}) yielding a discrete outcome $Q\in\mathbb{N}_0$. Let us thus begin by defining the space of zero-mean square-summable functions on $\mathbb{N}_0$, that is, square-summable with respect to the measure $f^{(0)}=(f^{(0)}_q) =\int H(q;z)F_0(dz)$ representing the intensity at the reference parameter $\boldsymbol{\theta}_0$,
\begin{equation}
\begin{aligned}
    \ell^2_0(f^{(0)})=\bigg\{&g:\mathbb{N}_0\rightarrow\mathbb{R}~\big\lvert ~\sum_{q=0}^{\infty}g(q) f^{(0)}_q=0,\\
    &\sum_{q=0}^{\infty}\abs{g(q)}^2 f^{(0)}_q<\infty\bigg\}.
\end{aligned}
\end{equation}
As in Appendix \ref{AppendixDIIFefficiency} we will assume that the support of $F_0$ is an infinite set contained in $A=[-a,a]\subset\mathbb{R}$, where $a\in(0,z_R)$ is such that $a =\sup
\{z~|~ z\in {\rm support}(F_0)\}$.

Again let us define the statistical model as the collection of distributions that are dominated by the reference distribution, $\mathcal{P}=\{F:F \ll F_0\}$, and consider the map from a neighborhood of $t=0$ in $[0,\infty)$ to the model $\mathcal{P}$, given by Eq.~(\ref{onepiece}) with the exponential family (\ref{normafamily}). Given the image intensity $f^{(t)}_q=\int H(q;z)F_t(dz)$, we can write the score functions as 
\begin{equation}
    S_u(q)=\left.\frac{\partial}{\partial t}\right\vert_{t=0}\log f_q^{(t)}=\frac{\int H(q;z)u(z)F_0(dz)}{f_q^{(0)}}.
\end{equation}
Recall that the function $H$ for the SPADE model is given by 
\begin{eqnarray}
    H(q;z)=\frac{4z_R^2z^{2q}}{(z^2+4z_R^2)^{q+1}}.
\end{eqnarray}
Like in Appendix~\ref{AppendixDIIFefficiency}, let us define the space of score functions as $\mathfrak{S}$. We aim to show that the co-tangent space  $\mathcal{T}\equiv\text{span}\{\mathfrak{S}\}$ is dense in the Hilbert space $\ell^2_0(f^{(0)})$ of zero-mean square-summable functions, that is, we aim to show:
\begin{eqnarray}
    \overline{\mathcal{T}}=\ell^2_0(f^{(0)}).
\end{eqnarray}
We will do this once more by showing that $\mathcal{T}^\perp=\{0\}$, which implies that an IF belonging to $\ell^2_0(f^{(0)})$ is automatically efficient.

Let us consider a function $g\in \mathcal{T}^{\perp}$ and take its inner product with $S_u\in \mathcal{T}$:
\begin{equation}
\begin{aligned}
\label{SPADEinnerProduct}
    0&=\langle g,S_u \rangle_{F_{0}}=\sum_{q=0}^{\infty}g(q)S_u(q)f_q^{(0)}\\
    &=\sum_{q=0}^{\infty}g(q)\int_A H(q;z)u(z)F_0(dz)
\end{aligned}
\end{equation}
We wish to swap the sum and the integral. By Tonelli's theorem this can be done only if $g(q)H(q;z)u(z)$ is absolutely integrable with respect to the measure $f_q^{(0)}\times F_0(dz)$ i.e.,
\begin{eqnarray}
    \sum_{q=0}^{\infty} \int_A \abs{g(q)} \abs{H(q;z)} \abs{u(z)} F_0(dz)<\infty.
\end{eqnarray}
Since $u(z)\in L^{\infty}(F_0)$, we have $|u(z)|\leq \|u\|_{\infty}$ for all $z\in A$. Moreover since $H(q;z)$ is a real and positive function one can use Cauchy-Schwartz to bound the LHS above as
\begin{eqnarray}
    &&\|u\|_{\infty}\sum_{q=0}^{\infty} \abs{g(q)} \int_A  H(q;z) F_0(dz) \nonumber \\
    &=& \|u\|_{\infty}\left(\sum_{q=0}^{\infty}\abs{g(q)}^2f^{(0)}_q\right)^{\frac{1}{2}}\left(\sum_{q=0}^{\infty}f^{(0)}_q\right)^{\frac{1}{2}}\\
    &<&\infty. \nonumber
\end{eqnarray}
As such Eq.~(\ref{SPADEinnerProduct}) can be written as
\begin{eqnarray}
    \int_A G(z) u (z) F_0(dz)=0,
\end{eqnarray}
where 
\begin{eqnarray}
    G(z)\equiv \sum_{q=0}^{\infty}g(q)H(q;z).
\end{eqnarray}
Since $u(z)$ is an arbitrary zero-mean measurable function with respect to $F_0$, and $G$ is zero-mean, this implies 
that 
$G(z)=0,\,\,F_0-a.e. \, z\in\mathbb{R}$.
As before, in order to show that $G(z)$ 
is identically zero, it suffices to prove that $G(z)$ is real analytic on $A$. Let us define $\xi(z)\equiv z^2/(z^2+4z_R^2)$ such that
\begin{eqnarray}
    0\leq \xi(z)\leq \frac{a^2}{a^2+4z_R^2}\equiv\xi_{\text{max}}<1.
\end{eqnarray}
As such we have
\begin{equation}
    H(q;z)=(1-\xi(z))\xi(z)^q,
\end{equation}
and 
\begin{equation}
G(z)=(1-\xi(z))\sum_{q=0}^{\infty}g(q)\xi(z)^q=(1-\xi(z))P(\xi(z)),
\end{equation}
where we have defined $P(\xi)$ as the power series
\begin{eqnarray}
\label{PowerSeries}
    P(\xi)=\sum_{q=0}^{\infty}g(q)\xi(z)^q.
\end{eqnarray}
In order to prove that $G(z)$ is real analytic on $A$ we will prove that $P(\xi)$ converges absolutely on $A$. To do this, we will note that since $F_0$ is not a point mass at $z=0$, and $a$ has been chosen as the supremum of (the absolute value of) the support of  $F_0$, any $ \xi_*\in(0,\xi_{\text{max}})$ satisfies $F_0(B)>0$ where $B=\{z\in A:\xi(z)\geq \xi_*\}$. As such given that $H(q;z)$ is positive, we can bound the reference intensity from below as follows,
\begin{equation}
\begin{aligned}
    f^{(0)}_q&=\int_A H(q;z) F_0(dz)\geq\int_B H(q;z) F_0(dz)\\
    &=\int_B (1-\xi(z))\xi(z)^q F_0(dz)\geq c\xi_{*}^q.
\end{aligned}
\end{equation}
With this bound and using the $\ell^2$ square-summability of $g$ we can then write 
\begin{eqnarray}
    c\sum_{q=0}^{\infty}\abs{g(q)}^2\xi_{*}^q\leq\sum_{q=0}^{\infty}\abs{g(q)}^2f^{(0)}_q<\infty.
\end{eqnarray}
Given that for $\xi_*$, the series on the LHS converges, by the root test we can conclude that the radius of convergence of this series is
\begin{eqnarray}
    R=\frac{1}{\limsup_{n\rightarrow\infty}\abs{g(q)}^{\frac{2}{q}}}>\xi_*.
\end{eqnarray}
This implies that the power series given by Eq.~(\ref{PowerSeries}) has convergence radius $R^\prime$ at least $\sqrt{\xi_*}$.
Since $\xi_*<\xi_{\rm max}$ was chosen arbitrarily, we can impose the condition 
\begin{eqnarray}
    \xi_*>\xi_{\text{max}}^2\equiv\left(\frac{a^2}{a^2+4z_R^2}\right)^2<1.
\end{eqnarray}
When this condition is satisfied we have $R^\prime >\xi_{\text{max}}$ and thus the power series (\ref{PowerSeries}) converges for all $\xi\in(0,\xi_{\text{max}})$. A power series is analytic on the interior of its radius of convergence and so $P(\xi)$ is analytic $\forall\xi\in(0,\xi_{\text{max}})$ and thus implicitly real analytic $\forall z\in A$. Since $\xi(z)=z^2/(z^2+4z_R^2)$ is also real analytic on $A$, so too is $G(z)$.

We can now use the identity theorem for real analytic functions to prove that if $G(z)$ is real analytic in $A$ and $G(z)=0,\,\,F_0-a.e. \, z\in A$, then $G(z)=0$ for all $ z\in A$. We will not explicitly write out the proof as it is identical to the one in Appendix \ref{AppendixDIIFefficiency}. 
As such, since $G(z)=\left(1-\xi(z)\right)P(\xi)=0\,\,\forall z\in A$, this means that $P(\xi)=0\,\,\forall z\in A$. Since the terms of a power series are linearly independent we conclude that $g(q)=0\,\,\forall q\in\mathbb{N}_0$. Seeing that the only element in $\ell^2_{0}(f^{(0)})$ orthogonal to the co-tangent space $\mathcal{T}$ is the zero vector, we conclude that its orthogonal complement $\mathcal{T}^{\perp}$ is trivial and thus $\mathcal{T}$ is dense in $\ell^2_{0}(f^{(0)})$. In conclusion any element in $\ell^2_{0}(f^{(0)})$ can be arbitrarily well approximated by elements in the space of scores $S$ and thus an IF $\varphi\in \ell^2_{0}(f^{(0)})$ implies that $\varphi$ is automatically efficient.

\bibliography{bibliography}

\end{document}